\documentclass[aps,a4paper,twocolumn,superscriptaddress,groupedaddress]{revtex4}  
\usepackage{graphicx}  
\usepackage{dcolumn}   
\usepackage{bm}        
\usepackage{amssymb}   

\usepackage{amsmath,amssymb,amsfonts,amsthm,bbm} 
\usepackage{graphicx}
\usepackage{tikz}
\usepackage{color}
\usepackage{verbatim}
\usepackage{layout} 
\usepackage{mathrsfs}  
\usepackage{import}
\usepackage{psfrag}
\usepackage{pstricks,pst-node}

\newcommand{\Z}{\mathbb{Z}}

\newcommand{\dd}{\mathrm{d}}

\newcommand{\ed}{\mathrm{e}}

\definecolor{golden}{rgb}{0.9,0.7,0.2}
\definecolor{pink}{rgb}{1,0.5,0.5}

\newcommand{\caC}{{\mathcal C}}

\newcommand{\caE}{{\mathcal E}}
\newcommand{\caF}{{\mathcal F}}

\newcommand{\caH}{{\mathcal H}}

\newcommand{\coupling}{{\mathsf J}}  
\newcommand{\NumLevels}{{\mathsf N}}
\newcommand{\class}{{\mathsf c}}
\newcommand{\str}{ |}
\newcommand{\norm}{ \|}
\newcommand{\beq}{\begin{equation}}
\newcommand{\eeq}{\end{equation}}

\newcommand{\wdr}{\textcolor{blue}}

\begin{document}

\title{{Scenario for delocalization in translation invariant systems}}
\author{Wojciech De Roeck} 
\affiliation{\footnotesize{Instituut voor Theoretische Fysica, KU Leuven, Belgium, \texttt{Wojciech.DeRoeck@fys.kuleuven.be}}}
\author{Fran\c cois Huveneers}
\affiliation{\footnotesize{CEREMADE, Universit\' e Paris-Dauphine, France, \texttt{huveneers@ceremade.dauphine.fr}}}

\date{\today}
\begin{abstract}
\noindent
We investigate the possibility of Many-Body Localization in translation invariant Hamiltonian systems, which was recently brought up by several authors. 
A key feature of Many-Body Localized disordered systems is recovered, namely the fact that resonant spots are rare and far-between. 
However, we point out that resonant spots are mobile, unlike in models with strong quenched disorder, 
and that {these mobile spots constitute a possible mechanism for delocalization}, albeit possibly only on very long timescales. 
In some models, this argument for delocalization can be made very explicit in first order of perturbation theory in the hopping. 
For models where this does not work, we present instead a non-perturbative argument that relies solely on ergodicity inside the resonant spots. 

\end{abstract}

\maketitle

\section{Introduction}\label{section: Introduction}

The theory of Many-Body Localization {(MBL)} is being shaped and sharpened  right now.  Briefly said, MBL is a phase of matter in which equilibrium statistical mechanics does not apply: there is no thermalization and no transport, see \cite{husereview}. For a many-body system consisting of non-interacting fermions in a disordered potential, MBL is an easy consequence of the fact that one-fermion wave functions are Anderson localized \cite{andersonabsence}. 
Whereas the first systematic treatment of MBL appears in \cite{BaskoAleinerAltshuler}, in the context of interacting electrons in a disordered potential, recent numerical and theoretical work on disordered spin chains \cite{OganesyanHuse,PalHuse}, 
the contrast with the ergodic properties of eigenfunctions conjectured for `non-localized' systems (ETH: Eigenstate Thermalization Hypothesis, see \cite{srednicki,deutsch}) 
and the connection to dynamical phase transitions \cite{Polkovnikovetal} have added a lot of appeal to the subject. 
Recently, a mathematically rigorous underpinning of the phenomenon has been provided as well \cite{Imbrie}.    

Whereas most of these considerations concern quenched disorder,
it has recently been suggested that also thermal (or configurational) disorder could serve the same purpose and localize a system
\cite{KaganMaksimov,Carleoetal,GroverFisher,DeRoeckHuveneers,SchiulazMueller,Hickey}. In other words, it has been suggested that MBL can occur also in systems where the Hamiltonian has no random terms and is translation-invariant. In such a scenario, the `effective' randomness is provided by the initial state. 
To discuss this issue, we introduce in Section \ref{section: First model} a lattice quantum model of interacting bosons.  This model is very similar to the Bose-Hubbard model, and it contains the latter as a special case, but we keep the discussion general to highlight the basic mechanism at work. 
At high energy density (high temperature) and small hopping, resonant spots appear to be as rare as in some quenched disordered systems. 
This is the key observation leading to the conjecture that an {MBL} phase {exists} in translation invariant systems. 

For systems having all their eigenstates localized,
a description of the localized phase in terms of local conserved quantities was proposed in \cite{HuseOganesyan}. Following \cite{HuseOganesyan}, we refer to this case as `full MBL'. 
Such a description does not carry over for the putative localized phase of our system, as arbitrarily large resonant regions may appear everywhere in the system.
This is related to the fact that certainly not all eigenstates can be localized in our translation-invariant systems; in particular, at low energy density, we expect an ergodic phase.  
Nevertheless, in  Section \ref{sec: many body localized phase}, we provide a {hypothetical} characterization that takes this issue explicitly into account,  
while still retaining most of the features of a localized phase, like absence of thermalization and vanishing transport coefficients.

The main distinction between (quenched) disordered and translation invariant systems 
shows up when considering the effect of resonances.  The key question is whether they can delocalize the system. 
For quenched disordered Hamiltonians {where full MBL is expected} ({at strong disorder}), like in \cite{Imbrie,OganesyanHuse,PalHuse}, 
the problem trivializes:
resonant spots form small, isolated islands in physical space, their location determined by regions of anomalous disorder realization, and, therefore, they produce no transport. 
When translation invariance is restored, resonant spots become possibly mobile, exactly because they are not tied to particular regions. 
If this possibility {is} realized, the resonant spots, also called `ergodic spots' later, 
could act as carriers of energy (or any other conserved quantity, for that matter) and they {could} thus delocalize the system. 
We refer to this scenario as `percolation in configuration space' 
since it corresponds to the case where all states connected via resonant transitions form a giant cluster in configuration space. 

We proceed to a detailed analysis in a few different instances of {the} model {introduced in Section \ref{section: First model}, 
varying dimension and the precise form of the hopping term.
In several instances, we find mobile ergodic spots already in first order in the hopping. 
In other cases, no percolation is observed at first order. 
We then develop a non-perturbative argument to show that percolation in configuration space does occur
(an argument of a similar flavor was developed by \cite{HuseNandkishore}). 
Though expressed in a particular set-up, the reasoning is very general, as it only relies on an ergodicity assumption inside the ergodic spots.
We believe that it applies to all generic translation invariant lattice Hamiltonian with short range interaction.

\emph{Acknowledgements.}
This article has benefitted a lot from comments and suggestions by David Huse, at an early stage of the work.  We are also grateful to 
John Imbrie, Markus M\"uller, and Rahul Nandkishore for sharing and explaining their work, and pointing out several issues related to this article. 
The authors thank BIRS (Banff, workshop 13w508),  IAS (Princeton), Rutgers University,  KU Leuven, PCTS (Princeton) and Universit\'e Paris-Dauphine, for hospitality and financial support.
W.D.R also thanks the DFG (German Research Fund) and the Belgian Interuniversity Attraction Pole  (P07/18 Dygest) for financial support.

\section{The model}\label{section: First model}

We introduce a quantum lattice system in a large volume $V \subset \Z^d$, and study it in the thermodynamic limit $V\to \infty$.
We will often restrict ourselves to $d=1$.
Although we  work with a rather abstract model in order to showcase the dominant features, 
the considerations developed here apply equally well to more realistic Hamiltonians \cite{SchiulazMueller,DeRoeckHuveneers}.  
For concreteness, we adopt a vocabulary that is close to the Bose-Hubbard model in \cite{DeRoeckHuveneers}, 
and we think of each lattice site $x \in V$ as containing a variable number of bosons $\eta_x\in \{ 0, \dots , \NumLevels\}$, where $\NumLevels$ is a cutoff on the occupation number per site. 
Consequently, we have a preferred product basis in the many-body Hilbert space,  consisting of \emph{classical configurations}  $|\eta \rangle = |(\eta_x)_{x\in V}\rangle$. 

The bosons  interact  locally at each site and we assign to each (local) occupation number $n$ an energy $\mathcal E(n)$, such that  $0=\mathcal E(0) < \mathcal E(1) \dots < \mathcal E(\NumLevels)$. 
The Hamiltonian is of the form    
\begin{equation}\label{general form Hamiltonian}
H 
\; = \; 
E^{(0)} \, + \, \coupling U
\; = \; 
\sum_{x\in V} \big( E^{(0)}_x \, + \, \coupling U_x \big).
\end{equation}
$E^{(0)}$ is a diagonal matrix in the $\{ | \eta \rangle \}$ basis, taking account of the interaction between bosons, 
while $U$ allows for hopping
\begin{equation}\label{original Hamiltonian}
\langle \eta | E_x^{(0)}| \eta \rangle \; = \;  \mathcal E({\eta_x}), 
\quad 
U_x 
\; = \; 
\frac{1}{2d}\sum_{y\sim x} \big( b_{x}^* b_{y} + \text{h.c.} \big),
\end{equation}
where $b_x$ and $b_x^*$ are bosonic annihilation/creation operators with a cutoff: 
\begin{equation*}
b_x | \dots , \eta_x , \dots \rangle \; = \; \sqrt{\eta_x} \, | \dots, \eta_{x}-1, \dots \rangle \quad \text{if} \quad \eta_x \ne 0,
\end{equation*}
and $ b_x | \eta \rangle \; = \; 0$ otherwise. 
{The specific form in \eqref{original Hamiltonian} is examplary: more generic terms for the interaction and the (short range) hopping can and will be considered.}

{We choose units such that the highest on-site energy $\caE(\NumLevels) $ is of order $1$ and we treat $\coupling$ as a dimensionless perturbative parameter.  The smallest on-site energy spacing will be of order $1/\NumLevels$ and we assume to be in the regime of strong interactions compared to the hopping: 
\begin{equation}\label{strong interaction regime}
0 \; \le \; \coupling\NumLevels \; \ll \; \frac{1}{\NumLevels} \; \ll \; 1. 
\end{equation}
For $\coupling = 0$, the eigenstates of $H$ are the classical configurations $\str \eta \rangle$. They are perfectly localized in physical space 
(a more precise definition of localization will be given in Section \ref{sec: many body localized phase}).}

\subsection{Rare resonant spots}\label{sec: rare resonant spots} 
We choose the characteristics of our system so as to make the analogy with (quenched) disordered systems as perfect as possible in the regime $\coupling \to 0$.
In the absence of any external disorder, randomness manifests itself in the system via the initial state, or, equivalently, when considering a thermal ensemble. 
We will often say that something is true for a typical configuration $\eta$, 
and this hence refers to the natural counting (i.e.\ with equal weights) of configurations as discrete elements in $\{0,\ldots, \NumLevels\}^{V}$. 
Alternatively, one can think of this as the infinite-temperature ensemble, in which $\eta_x$, $x \in V$, are i.i.d.\ random variables.  
In any case, considering the ensemble of configurations should mimic the case of quenched independent disorder on each site. 

We assume the interaction between particles to be strongly anharmonic (nonlinear):
A harmonic (linear) interaction would mean that $\caE(n)$ is linear in $n$.  
To have a maximally anharmonic interaction, 
we imagine  $\mathcal E(1), \dots , \mathcal E(\NumLevels)$  to be  a typical realization of a process that throws $\NumLevels$ points at random on an interval with length of order $1$.
Since the set of values $\caE(n)$ is given by the same realization at all sites, the model is still translation invariant.  Choosing $\caE$ in this manner, we make \emph{resonances} as rare as possible, as we explain now. 

In general, we say that two configurations $ \eta $ and $ \eta'$ are resonant in first order in $\coupling$ if 
\beq\label{def: first order resonance}
\str \langle \eta | E^{(0)} | \eta \rangle -  \langle \eta' | E^{(0)} | \eta' \rangle  \str  \;  \ll  \;    \coupling  \str     \langle \eta | U | \eta' \rangle \str,
\eeq
see also Section \ref{subsection: Basic picture RG} for a motivation of this definition. 
Because of our choice of $\caE$, and inequality \eqref{strong interaction regime}, we can simplify this: two configurations $ \eta $ and $ \eta'$ are resonant (in first order)
if they have the same interaction energy $E^{(0)}$ and they are connected in first order by the perturbation $U$:
\begin{multline}\label{First order resonance}
\langle \eta' | E^{(0)} | \eta' \rangle \; = \; \langle \eta | E^{(0)} | \eta \rangle
\qquad \text{and} \\[1mm]
\langle \eta' | U_x | \eta \rangle \; \ne \; 0 
\quad \text{for some} \quad 
x\in V.
\end{multline}
If the interaction were harmonic, then $\mathcal E(\NumLevels) - \mathcal E({\NumLevels-1}) = \dots = \mathcal E(1) - \mathcal E(0)$ such that any first order transition is resonant: $
\langle \eta' | U_x | \eta \rangle \ne 0$ {for some  $x \in V$ implies}  $\langle \eta' | E^{(0)} | \eta' \rangle  =   \langle \eta | E^{(0)} | \eta \rangle$.
{(see the left panel on figure \ref{figure: harmonic and anharmonic}).}
However, for our anharmonic model, first order resonances are rare for large $\NumLevels$.  
In what follows, let us  restrict ourselves  to $d=1$ for notational convenience. 
We find that two configurations $\eta,\eta'$ are resonant if and only if there is some site $x \in V $  such that 
$\eta_y=\eta'_{y}$ for all $y \neq x,x+1$ and one of the following two conditions holds
\begin{multline*}
\eta_x'    \; = \;  \eta_{x+1} \; = \; \eta_x + 1 \; = \;  \eta_{x+1}' +1 
\qquad \text{or} \\[1mm]
\eta_x'    \; = \;  \eta_{x+1} \; = \; \eta_x - 1 \; = \;  \eta_{x+1}' -1.  
\end{multline*}
This is illustrated  {on the right panel of} figure \ref{figure: harmonic and anharmonic}. 
We call such a bond (pair of adjacent sites) $(x,x+1)$ a resonant bond (spot) for the configuration $\eta$ (or $\eta'$).   
The important observation here is that, for a typical configuration $\eta$, the resonant spots are typically isolated and rare, the distances between them being of order  $\NumLevels$.

\begin{figure}[h!]
\begin{tikzpicture}[xscale=0.58,yscale=0.8]


\draw [>=stealth,->] (-0.2,0) -- (5.2,0);
\draw [>=stealth,->] (0,-0.2) -- (0,3.2);

\draw (4.9,-0.1) node[below]{$x$} ;
\draw (0,2.9) node[left]{$\mathcal E(\eta_x)$} ;

\draw [>=stealth,->] (6.8,0) -- (12.2,0);
\draw [>=stealth,->] (7,-0.2) -- (7,3.2);

\draw (11.9,-0.1) node[below]{$x$} ;
\draw (7,2.9) node[left]{$\mathcal E(\eta_x)$} ;


\draw [ultra thick] (0.1,1) -- (0.9,1);
\draw [ultra thick] (1.1,2) -- (1.9,2);
\draw [ultra thick] (3.1,1.5) -- (3.9,1.5);
\draw [ultra thick] (4.1,1) -- (4.9,1);

\draw [ultra thick,dashed] (0.1,0.5) -- (0.9,0.5);
\draw [ultra thick,dashed] (1.1,2.5) -- (1.9,2.5);
\draw [ultra thick,dashed] (3.1,1) -- (3.9,1);
\draw [ultra thick,dashed] (4.1,1.5) -- (4.9,1.5);

\draw [>=stealth,->] (0.5,1) -- (0.5,0.6);
\draw [>=stealth,->] (1.5,2) -- (1.5,2.4);
\draw [>=stealth,->] (3.5,1.5) -- (3.5,1.1);
\draw [>=stealth,->] (4.5,1) -- (4.5,1.4);

\draw [ultra thick] (-0.1,0.5) -- (0,0.5);
\draw [ultra thick] (-0.1,1) -- (0,1);
\draw [ultra thick] (-0.1,1.5) -- (0,1.5);
\draw [ultra thick] (-0.1,2) -- (0,2);
\draw [ultra thick] (-0.1,2.5) -- (0,2.5);


\draw [ultra thick] (7.1,0.95) -- (7.9,0.95);
\draw [ultra thick] (8.1,2.04) -- (8.9,2.04);
\draw [ultra thick] (10.1,1.3) -- (10.9,1.3);
\draw [ultra thick] (11.1,0.95) -- (11.9,0.95);

\draw [ultra thick,dashed] (7.1,0.1) -- (7.9,0.1);
\draw [ultra thick,dashed] (8.1,2.53) -- (8.9,2.53);
\draw [ultra thick,dashed] (10.1,0.95) -- (10.9,0.95);
\draw [ultra thick,dashed] (11.1,1.3) -- (11.9,1.3);

\draw [>=stealth,->] (7.5,0.95) -- (7.5,0.2);
\draw [>=stealth,->] (8.5,2.04) -- (8.5,2.43);
\draw [>=stealth,->] (10.5,1.3) -- (10.5,1.05);
\draw [>=stealth,->] (11.5,0.95) -- (11.5,1.2);

\draw [ultra thick] (6.9,0.1) -- (7,0.1);
\draw [ultra thick] (6.9,0.95) -- (7,0.95);
\draw [ultra thick] (6.9,1.3) -- (7,1.3);
\draw [ultra thick] (6.9,2.04) -- (7,2.04);
\draw [ultra thick] (6.9,2.53) -- (7,2.53);

\end{tikzpicture}

\caption{
\label{figure: harmonic and anharmonic}
First order hopping in $\coupling$ for harmonic and anharmonic interactions in $d=1$. 
On the left, interaction is harmonic: hopping never results in frequency mismatches.
On the right, interaction is anahormonic: resonances only occur when two levels are swapped (most right interaction).  
}
\end{figure}
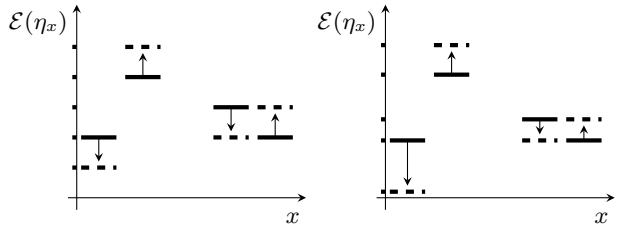

We conclude that our system has the same basic features as a disordered Hamiltonian, 
where the on-site disorder takes values in a discrete set of $\NumLevels+1$ elements. 
The maximal number of particles per site, $\NumLevels$, serves thus as a control parameter on the density of resonant spots, which is of order $1/\NumLevels$. 

Finally, let us stress that taking $\caE$ random is a way to implement anharmonicity, but it is not a necessity.  In fact, a simple choice like $\mathcal E(n) = (n/\NumLevels)^2$ would be {perfectly suited} as well. 
In that case, the Hamiltonian (\ref{general form Hamiltonian}-\ref{original Hamiltonian}) is the Bose-Hubbard Hamiltonian, up to the cutoff in occupation number and rescaling of parameters. 

\section{A priori restrictions on localization} \label{sec: many body localized phase}

There are some basic intuitive obstructions to a tentative localized phase for the system introduced above:
the existence of ergodic states at all energy densities and the translation invariance of the Hamiltonian.   
We discuss them and then we propose a description of the {hypothetical} localized phase that takes these objections into account. 

\subsection{No uniform localization}
In  models with strong quenched disorder and a finite dimensional on-site space, as considered in \cite{Imbrie,OganesyanHuse,PalHuse}, 
the MBL phase can be characterized by saying that all eigenstates of the Hamiltonian are, 
in some sense, close to the eigenstates of the unperturbed system ($\coupling=0$), i.e.\ to the classical configurations $| \eta \rangle$. 
This is the regime of full MBL.
Based on this, a description in terms of a complete set of conserved quantities was proposed in \cite{HuseOganesyan}.

This picture cannot survive in our model, since the behavior of the system will inevitably depend on the initial state. 
Indeed, for example, under the additional restriction that the particle density is much smaller than $1$ (the energy density will then also be much smaller than $1$), most of the particles are typically isolated and do not interact.  
In this regime, the comparison with a disordered system breaks down, and there is no reason to expect anything but normal transport and thermalization. 
In fact, in the zero-density limit, as boson-boson scattering becomes negligible, the transport can even become ballistic (a ballistic configuration is depticted in figure \ref{figure: coexistence perconlating non percolating} as $\eta_{(1)}$).

Moreover, ergodic behavior cannot be suppressed by just imposing a high enough particle density, as indeed, 
in a typical configuration there will inevitably be regions where the local density is very low (a large deviation).
In fact, we can actually expect fully delocalized eigenstates at any density of particles (and thus also at any energy density),
originating from `flat' configurations like the bottom one in figure \ref{figure: coexistence perconlating non percolating}, 
but now not necessarily with zero occupation, i.e.\ $\eta_x=n$ for all $x$ and some occupation number $n$. 

Such ergodic states occupying the full volume $V$ become quickly exceptional in the thermodynamic limit $V \rightarrow \infty$.
Therefore, they do not need to have any impact on the transport properties, but their existence rules out a characterization based on a complete set of local conserved quantities.
This situation is quite analogous to quenched disordered systems at sufficiently low disorder,
where most authors expect a localization-delocalization transition in function of the energy density \cite{BaskoAleinerAltshuler,Pollmann}  (see however \footnote{We question this, exactly because of the analogy to the situation in the present paper, and we \cite{De Roeck Huveneers Mueller Schiulaz} are currently investigating whether ergodic spots can destroy the localized phase in quenched disorder systems with non-full MBL.}).

On the other hand, regions with divergent localization length also occur in {full MBL} systems, as in \cite{Imbrie,OganesyanHuse,PalHuse}.
In that case, however, the location of these regions is determined by the realization of disorder (large deviation of the disorder). 
In contrast, for translation-invariant systems, such regions appear everywhere, their location depending on the state in Hilbert space.

\subsection{Translation invariance and symmetries}
For periodic boundary conditions, the Hamiltonian defined by (\ref{general form Hamiltonian}-\ref{original Hamiltonian}) is translation-invariant. 
Therefore, as observed by \cite{SchiulazMueller}, if one reasonably assumes that there are no degeneracies in the spectrum, the true eigenstates must be translation invariant as well, contradicting any notion of genuine localization. 
Nevertheless, strict translation invariance can be broken by another choice of boundary conditions. 
MBL in our system amounts then to spontaneous symmetry breaking of the translation-invariance, 
in complete analogy with the classical spontaneous symmetry breaking of a local order parameter by a boundary field.  

Additionally, several discrete symmetries, such as rotations or reflections, can leave the Hamiltonian invariant.
Since it is easy to break them in a robust way, i.e.\@ independently of boundary conditions, by introducing an additional interaction term at each site, 
we will not further consider them here.

\subsection{Description of the hypothetical MBL phase}\label{sec: description of putative mbl}

Let us consider a Hamiltonian of the type (\ref{general form Hamiltonian}-\ref{original Hamiltonian}). 
We assume that all geometrical symmetries are broken by boundary conditions and possibly additional interaction terms.  
Let $\Omega$ be \wdr{a} unitary change of basis that diagonalizes the Hamiltonian $H$ in the $\str \eta \rangle$-basis,
\begin{equation}\label{eq: classical hamiltonian}
H=  \Omega H_{free} \Omega^*, \qquad   H_{free}= H_{free}(\eta). 
\end{equation} 
Consider  now a local  operator $O'_x$ acting on a small spatial set containing a given point $x\in V$ and expand
\begin{equation}\label{MBL characterization 1}
\Omega^*  \, O'_x \, \Omega \; = \; \sum_{A \ni x} O_A,
\end{equation} 
where the sum runs over all connected subsets $A \subset V$ containing $x$, and where $O_A$ is an operator acting locally in the set $A$.
For {full MBL systems}, 
localization amounts to the statement {that $\Omega$ can be chosen such} that the action of $O_A$ on any state produces an exponentially small factor $\ed^{-c |A|}$, except in rare resonant regions. 
Another way to say this is that the operator norm of $O_A$ decays exponentially with $|A|$, or, that $\Omega$ acts quasilocally, except in rare resonant regions. 

In the translation invariant case, however, localization means that the operator $O_A$  decays exponentially when acting on typical states in $A$ 
(but for example not on states with an anomalously low density of particles in $A$, states that are exceptional). 
A possible way of making this precise is to consider the Hilbert-Schmidt norm of $O_A$, instead of the operator norm: 
\begin{equation}\label{MBL characterization 2}
\mathrm{tr} \, (O_A^* O_A) \; \sim \; \ed^{-c |A|},
\end{equation}
with $\mathrm{tr}(\cdot)= \frac{1}{\mathrm{dim}} \mathrm{Tr}(\cdot) $ the normalized trace.  The average over states that is present in `$\mathrm{tr}(\cdot)$' eliminates the exceptional states.  

The above discussion applies in particular to the Hamiltonian $H_{free}$, since $H$ was written as a sum of local operators: $H = \sum_x H_x$.
{Let us write correspondingly $H_{free}(\eta)$ as $H_{free}(\eta) = \sum_x H_{free,x}(\eta)$.
Each $H_{free,x}(\eta)$ is a sum local functions $f_A(\eta_A)$ with $A$ centered on $x$, cfr.\ \eqref{MBL characterization 1}. 
For fully MBL systems, $\norm f_A \norm_{\infty}$ decays with $\str A\str$,  but the decay depends on the site $x$ (via the local disorder realization), 
whereas for translation invariant systems in the hypothetical localized phase, the decay is uniform in $x$, but dependent on the configuration $\eta$ around $x$. 
(it becomes arbitrarily slow for exceptional configurations $\eta$)}

The properties (\ref{MBL characterization 1}-\ref{MBL characterization 2}) suffice to derive physically meaningful information, 
such as the vanishing of transport coefficients at equilibrium or the breakdown of ETH. 
In \cite{DeRoeckHuveneers}, an approximate version of (\ref{MBL characterization 1}-\ref{MBL characterization 2}) is used to show that the thermal conductivity of 
a chain analogous to the Bose-Hubbard chain, decays faster than any power law as the temperature is sent to infinity.

\section{Resonances: quenched vs.\ thermal disorder}\label{section: RG picture and Resonances}

Recently,  an iterative scheme was proposed \cite{Imbrie,ImbrieSpencer} to  construct explicitly the change of basis $\Omega$ that diagonalizes $H$, for strongly disordered spin systems. 
This strategy  is very similar to the KAM scheme in classical mechanics where
the `localization' of some trajectories on submanifolds of the phase space is established through successive canonical transformations. 
As in these cases, the tendency to localization in our model is due to typical energy (frequency) mismatches.
{Here we first show, adopting the strategy of \cite{Imbrie,ImbrieSpencer}, how non-resonant transitions can be `removed'.
A non-perturbative analysis is necessary to understand the effect of resonances.
We next show why this question trivializes for systems where a full MBL phase is expected,
and why, a priori, it does not for the translation invariant system described by (\ref{general form Hamiltonian}-\ref{original Hamiltonian}), 
no matter how favorable the parameters $\coupling >0$ and $\NumLevels < + \infty$ are chosen.}

\subsection{Basic picture of the RG scheme}\label{subsection: Basic picture RG}
We follow \cite{Imbrie,ImbrieSpencer}. 
To find $\Omega$ such that $\Omega^* H \Omega$ is diagonal in the $\str\eta\rangle$-basis (recall Section \ref{sec: description of putative mbl}), 
we first try to determine perturbatively a change of basis $\tilde\Omega = \ed^{- \coupling \tilde A}$, for some anti-hermitian matrix $\tilde A$, 
such that $H' := \tilde\Omega^{*}H \tilde\Omega$ is now diagonal up to terms of order $\coupling^2$. 
If that works, the strategy can be iterated starting from $H'$ instead of $H$, with a coupling constant that is now $\coupling^2$ instead of $\coupling$. 
The scheme would thus converge very quickly as, after $n$ steps, the Hamiltonian would be diagonalized up to terms of order $\coupling^{2^n}$. 
This very naive picture will be considerably complicated by resonances. 

The first transformation $\tilde\Omega$ is obtained as follows. Assuming $\tilde A$ to be of order $1$, we expand in powers of $\coupling$:
\begin{eqnarray}
\tilde\Omega^{*}H \tilde\Omega 
&= &
\ed^{\coupling \tilde A} \big( E^{(0)} + \coupling U \big) \ed^{-\coupling \tilde A} \nonumber \\[1mm]
&= &
E^{(0)} \, + \, \coupling \big( U + [\tilde A,E^{(0)}] \big) \, + \, \mathcal O (\coupling^2). \label{first order perturabative expansion}
\end{eqnarray}
The first order in $\coupling$ vanishes if $\tilde A$ solves the equation $[E^{(0)}, \tilde A] = U$. 
Since $U = \sum_x U_x$, we can write $\tilde A = \sum_x \tilde A_x$, such that the equation $[E^{(0)}, \tilde A_x] = U_x$ is satisfied for every $x$:
\begin{equation}\label{A matrix at the first order}
\langle \eta' | \tilde A_x | \eta \rangle \; = \; \frac{\langle \eta' |U_x | \eta \rangle}{\langle \eta' |E^{(0)}|\eta' \rangle -  \langle \eta |E^{(0)} | \eta \rangle }  ,
\end{equation}
with the convention $0/0=0$ which means in particular that  $\langle \eta |\tilde  A_x | \eta \rangle = 0$ since the perturbation $U_x$ is off-diagonal. 
We see that  $\langle \eta' | \tilde A_x | \eta \rangle$ is well-defined provided that $ \eta, \eta'$ are not resonant, in the sense of \eqref{First order resonance} in Section \ref{sec: rare resonant spots}.   If we neglect those resonances, 
we would conclude that the perturbative expansion \eqref{first order perturabative expansion} is a posteriori justified. 
Moreover, a local observable rotated by $\tilde\Omega = \ed^{-\coupling  \tilde A}$ will stay local up to exponentially small corrections, since
\begin{equation}\label{Omega close to identity}
\tilde\Omega \; = \; \ed^{-\coupling \tilde A}, \quad \tilde A \; = \; \sum_{x\in V}  \tilde A_x, \quad \tilde A_x: \tilde A_x^* = -\tilde A_x,
\end{equation}
with $\tilde A_x$  local around $x$ and of order $1$, 
and in particular the perturbed eigenstates $\tilde \Omega \str \eta \rangle$ are similar to the classical configurations $\str \eta \rangle$. 

Let us now see how resonances affect this picture. 
We  split the interaction in two parts: 
\begin{equation*}
U \; = \; U_{res} + U_{per}, 
\end{equation*}
where $U_{res}$ collects all resonant transitions, and $U_{per}$ the rest.   More precisely, 
\begin{equation}\label{def: res interaction}
\langle \eta' | U_{per} | \eta \rangle:=\left\{ 
\begin{array}{ll}     
\langle \eta' | U | \eta \rangle    &  \text{for $\eta,\eta'$ non resonant,}  \\     0 &  \text{for $\eta,\eta'$ resonant.}       \end{array}\right. 
\end{equation}
We do the best we can: we solve only the equation $[E^{(0)},\tilde A] = U_{per}$ instead of the full $[E^{(0)},\tilde A] = U$. 
The matrix $\tilde A$ is now well defined, and $\tilde\Omega = \ed^{-\coupling \tilde A}$ is really of the type \eqref{Omega close to identity},
but we face the problem that we only obtain 
\beq \label{eq: renom ham}
H' = E^{(0)} + \coupling U_{res} + \mathcal O (\coupling^2).
\eeq
We thus need an extra, non-perturbative, step to get rid of the resonant coupling of order $\coupling$.  
In other words, we need to diagonalize the operator $H' = E^{(0)} + \coupling U_{res}$, which just amounts to diagonalizing $U_{res}$ inside blocks of constant  $E^{(0)} $. 
We will therefore refer to $U_{res}$ as the `resonant Hamiltonian'.

It is the nature of eigenstates of $U_{res}$  that eventually determines whether in first order the system is localized or not. 
Let $\tilde \Upsilon$ be {a} unitary {transformation} that diagonalizes $U_{res}$ in the $| \eta \rangle$ basis, 
such that the total change of basis (in the first step of the scheme) is $\tilde \Upsilon \tilde \Omega$ 
and the new perturbed eigenstates are given by $\tilde \Upsilon \tilde \Omega | \eta \rangle$.  
The main question is now whether {$\tilde\Upsilon$ can be chosen such} that most of these new eigenstates $\tilde \Upsilon \tilde \Omega | \eta \rangle$ are still close to the classical configurations $| \eta \rangle$ in most places.  
If the answer is `yes', 
also in later steps of  the scheme, then there is a strong case\footnote{
This however requires some thought. For example, there is the following issue that is absent in systems with quenched, smoothly distributed disorder. 
Since in higher orders, the hopping eventually becomes of range $\NumLevels$,
one needs to take into account the renormalization of the interaction by the hopping, as otherwise all transitions would seem resonant. This is explained in \cite{SchiulazMueller}.} for MBL in the sense of Section \ref{sec: many body localized phase}.
If instead the answer becomes `no' at some order, it is hard to {imagine} that higher orders could restore the localization. 
One is then led to the conclusion that the localized phase is absent.

\subsection{Resonances: systems with quenched disorder} \label{sec: resonances quenched}
For contrast, we first treat the case of  strongly quenched disordered systems, where it is simple to see why resonances do not induce any delocalization.
Let us consider as a standard example a one-dimensional spin-$1/2$ chain in a disordered field:
\begin{equation}\label{disordered spin chain}
H \; = \; \sum_{x\in V} \Big\{  \omega_x S_x^{3} \, + \, \coupling \,  (S^1_x  S^1_{x+1} +S^2_x ) \Big\}, 
\end{equation}
with $ S_x^{1},S_x^{2},S_x^{3}$ the usual Pauli matrices. 
We assume that $(\omega_x)_{x\in V}$ are i.i.d.\@ random variables and,
to make the connection with our model as direct as possible, 
we assume that the distribution of $\omega_x$ is concentrated on $\NumLevels+1$ values, 
that themselves look random, i.e.\ they are like the values $\caE(0), \ldots \caE({\NumLevels})$ introduced above.   
The classical configurations (eigenstates at $\coupling=0$) are in this model $\str (\eta_x)_{x\in V} \rangle$ with $\eta_{x}=\pm 1, x \in V$, 
referring to the eigenstates of $S_x^{3}$  (spin up / spin down).  
A first order resonance between configurations $\eta$ and $\eta'$ occurs when, for some $x$, it holds that 
\begin{multline*}
\langle \eta' | S^1_x  S^1_{x+1} +S^2_x    | \eta \rangle \ne 0
\qquad \text{and} \\[1mm]
\omega_x (\eta_x - \eta'_{x}) + \omega_{x+1} (\eta_{x+1} - \eta'_{x+1}) \; = \; 0,
\end{multline*}
Since we assumed that the values $\caE(n)$ of $\omega_x$ are chosen in a generic way, this can only happen when  $\omega_x = \omega_{x+1}$. 

Note that the above definition of resonance is identical to that given in \eqref{First order resonance}, but now, for the sake of simplicity, we proceed differently:
We define the resonant  $U_{res}$ as 
$$U_{res} \; = \; \sum_{x:\, \omega_x = \omega_{x+1}} U_x,$$  
which slightly differs from the definition in \eqref{def: res interaction}; 
for example the configurations $\eta_{x,x+1}= (1,1), \eta_{x,x+1}=(-1,-1)$ on a bond with $\omega_x = \omega_{x+1}$ would not be resonant according to definition \eqref{def: res interaction},
but the interaction connecting them is included in $U_{res}$. 
The important point here is that we can characterize resonant bonds in a purely geometric way, independently of the configurations $\eta,\eta'$. 
For large $\NumLevels$, these bonds, i.e.\@ those satisfying $\omega_x = \omega_{x+1}$,  form small isolated clusters $\caC$, located at a typical distance $\NumLevels$ from each other. 
The absence of percolation {(in real space)}
of the clusters $\caC$ leads to localization, see also the left panel of figure \ref{figure: quenched and thermal disorder}. 
Indeed, the matrix $\tilde \Upsilon$ that diagonalizes $U_{res}$ takes the form 
\begin{equation}\label{tilde Psi disordered}
\tilde \Upsilon \; = \; \ed^{-\sum_{\mathcal C} B_{\mathcal C}},    \qquad B_{\mathcal C}^* = -B_{\mathcal C}
\end{equation}
with $
\mathcal C$ resonant clusters and  $B_{\mathcal C}$  acting within $\mathcal C$.
Since $B_{\mathcal C}$ and $B_{\mathcal C'}$ commute for $\mathcal C \ne \mathcal C'$, 
we see that $\tilde \Upsilon$ acts locally, and hence  the full change of  basis $\tilde \Upsilon \tilde \Omega$ obtained after the first renormalization step, is quasilocal, 
it rotates local operators into quasilocal ones and all perturbed eigenstates $\tilde \Upsilon \tilde \Omega\str \eta \rangle$ are similar to $| \eta \rangle$,
away from the clusters $\caC$, where they are locally delocalized.

\subsection{Resonances: translation invariant systems}\label{sec: Resonances: translation invariant systems}

In translation invariant systems, the above reasoning cannot be simply copied, and, as we will see in Section \ref{section: Percolation resonances Model I},  its conclusion could be wrong. 

Consider  the graph ${\mathscr G}$ in configuration space that connects two classical configurations $\eta,\eta'$ if and only if they are resonant.  
The main, somehow surprising, point is that the connected components (classes) $\class$ of this graph could be very large even if for a typical configuration $\eta$, resonant spots are rare, 
see {the right panel of} figure \ref{figure: quenched and thermal disorder} for a hint. 
We refer to such behaviour as `percolation in configuration space' or simply `percolation of resonances', as opposed to percolation of resonant spots in real space.

\begin{figure}[h!]
\begin{tikzpicture}[scale=0.65]


\draw [very thin,color = gray!30] (0,0) grid  [step = 0.1] (3,3);
\draw (0,0) -- (3,0) -- (3,3) -- (0,3) -- (0,0);

\draw[thick,blue] (1,1) -- (1.1,1) -- (1.1,0.8);

\draw[thick,blue] (2.6,2.3) -- (2.6,2.2) -- (2.7,2.2);

\draw[thick,blue] (1.4,2.5) -- (1.5,2.5); 

\draw[thick,blue] (2.2,1.1) -- (2.2,1) -- (2.1,1);

\draw[thick,blue] (0.5,1.5) -- (0.5,1.6);

\draw[thick,blue] (2.8,0.3) -- (2.9,0.3) -- (2.9,0.4);

\draw[thick,blue] (1.5,1.8) -- (1.5,1.9);

\draw[thick, blue] (0.5,0.6) -- (0.6,0.6);

\draw[thick, blue] (0.2,2.7) -- (0.3,2.7);


\draw [>=stealth,->] (4.8,0) -- (11,0);
\draw [>=stealth,->] (5,-0.2) -- (5,3.3);

\draw (10.7,0) node[below]{$x$} ;
\draw (4.9,0.6) node[left]{$\eta$} ;
\draw (4.9,2.7) node[left]{$\eta'$} ;

\draw [thick,dotted] (5.1,0.625) -- (5.45,0.625); 
\draw [ultra thick]  (5.55, 0.5) -- (5.95,0.5);
\draw [ultra thick,blue]  (6.05, 1) -- (6.45,1);
\draw [ultra thick,blue]  (6.55, 0.75) -- (6.95,0.75);
\draw [ultra thick,blue]  (7.05, 0.75) -- (7.45,0.75);
\draw [ultra thick,blue]  (7.55, 0.5) -- (7.95,0.5);
\draw [ultra thick,blue]  (8.05, 0.25) -- (8.45,0.25);
\draw [ultra thick]  (8.55, 1) -- (8.95,1);
\draw [ultra thick]  (9.05, 0.5) -- (9.45,0.5);
\draw [ultra thick]  (9.55, 0.75) -- (9.95,0.75);
\draw [ultra thick]  (10.05, 0.25) -- (10.45,0.25);
\draw [thick,dotted]  (10.55, 0.625) -- (10.9,0.625);

\draw [thick,dotted] (5.1,2.625) -- (5.45,2.625); 
\draw [ultra thick,blue]  (5.55, 2.5) -- (5.95,2.5);
\draw [ultra thick,blue]  (6.05, 2.25) -- (6.45,2.25);
\draw [ultra thick,blue]  (6.55, 2.5) -- (6.95,2.5);
\draw [ultra thick,blue]  (7.05, 2.75) -- (7.45,2.75);
\draw [ultra thick,blue]  (7.55, 2.75) -- (7.95,2.75);
\draw [ultra thick,blue]  (8.05, 3) -- (8.45,3);
\draw [ultra thick,blue]  (8.55, 3) -- (8.95,3);
\draw [ultra thick]  (9.05, 2.5) -- (9.45,2.5);
\draw [ultra thick]  (9.55, 2.75) -- (9.95,2.75);
\draw [ultra thick]  (10.05, 2.25) -- (10.45,2.25);
\draw [thick,dotted]  (10.55, 2.625) -- (10.9,2.625);

\end{tikzpicture}

\caption{
\label{figure: quenched and thermal disorder}
First order resonances in $\coupling$ for quenched versus thermal disordered systems. 
Left panel: quenched disorder Hamiltonian in $d=2$. Resonances form fixed isolated non-percolating islands.
Right panel: translation invariant Hamiltonian in $d=1$, with an extra second neighbor interaction $b_x^* b_{x+2} + b_x b_{x+2}^*$. 
A bit of trial and error should convince the reader that it is possible to connect $\eta$ to $\eta'$ through a sequence of resonant transitions. 
The naive resonant spot in $\eta$ appears thus as part of a larger resonant cluster. 
}
\end{figure}
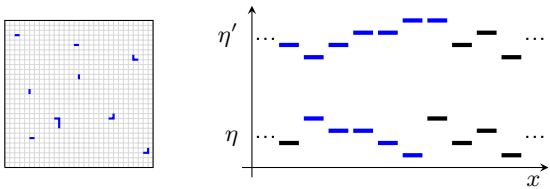

Since it is however not straightforward to talk about the size of the connected components \footnote{The number of connected components $\class$ cannot grow slower than $\str V\str^{\NumLevels}$ (as $\str V\str \to \infty$) due to the obvious constraint that the resonant Hamiltonian does not change the number of sites $x$  with $\eta_x=n$, for any $n \leq \NumLevels$. This polynomial constraint is however irrelevant for our question, as the number of vertices of ${\mathscr G}$ (number of configurations $\eta$) is exponential, namely $(\NumLevels+1)^{\str V \str}$.   }, we will define this phenomenon in a more pictorial way. 
First, we say that a site $x$ is frozen for a configuration $\eta$ if and only if 
$$
\eta'_x=\eta_x \qquad \text{for any $\eta' \in \class(\eta)$ (the class containing $\eta$)}
$$
This is a physically meaningful notion because $\class(\eta)$ is the set of configurations with which $\eta$ can hybridize (in first order) 
and hence a rotated state $\tilde \Upsilon \str\eta\rangle$ will be similar to $\eta$ on all frozen sites, but a priori not on the unfrozen sites.   
Note also that the unfrozen set depends just on the class $\class$, and not on $\eta \in \class$.
Now, we say that a class $\class$ has percolation in  configuration space if its unfrozen set  percolates in real space. 

At finite volume, both type of classes (percolating and not percolating) coexist, as illustrated in figure \ref{figure: coexistence perconlating non percolating}. 
As $V$ grows large, 
the number of configurations contained in either of the two classes determined by the examples in figure \ref{figure: coexistence perconlating non percolating} becomes quickly negligible.  
The real issue is then whether a typical state belongs to a class with or without percolation. 
In first order, the answer to this question appears to depend on detailed characteristics of the model,
while it becomes always `with percolation' at high enough order, as we show in the next section.

\begin{figure}[h!]
\begin{tikzpicture}[scale=0.65]

\draw [>=stealth,->] (-0.2,0) -- (11.2,0);
\draw [>=stealth,->] (0,-0.2) -- (0,3.5);

\draw (10.9,0) node[below]{$x$} ;
\draw (-0.1,0.5) node[left]{$\eta_{(1)}$} ;
\draw (-0.1,2) node[left]{$\eta_{(2)}$} ;

\draw [thick,dotted] (0.2,0.375) -- (0.45,0.375);
\draw [ultra thick] (0.55,0.25) -- (0.95,0.25);
\draw [ultra thick] (1.05,0.25) -- (1.45,0.25);
\draw [ultra thick] (1.55,0.25) -- (1.95,0.25);
\draw [ultra thick] (2.05,0.25) -- (2.45,0.25);
\draw [ultra thick] (2.55,0.25) -- (2.95,0.25);
\draw [ultra thick] (3.05,0.5) -- (3.45,0.5);
\draw [ultra thick] (3.55,0.25) -- (3.95,0.25);
\draw [ultra thick] (4.05,0.25) -- (4.45,0.25);
\draw [ultra thick] (4.55,0.25) -- (4.95,0.25);
\draw [ultra thick] (5.05,0.25) -- (5.45,0.25);
\draw [ultra thick] (5.55,0.25) -- (5.95,0.25);
\draw [ultra thick] (6.05,0.25) -- (6.45,0.25);
\draw [ultra thick] (6.55,0.25) -- (6.95,0.25);
\draw [ultra thick] (7.05,0.25) -- (7.45,0.25);
\draw [ultra thick] (7.55,0.25) -- (7.95,0.25);
\draw [ultra thick] (8.05,0.25) -- (8.45,0.25);
\draw [ultra thick] (8.55,0.25) -- (8.95,0.25);
\draw [ultra thick] (9.05,0.25) -- (9.45,0.25);
\draw [ultra thick] (9.55,0.25) -- (9.95,0.25);
\draw [ultra thick] (10.05,0.25) -- (10.45,0.25);
\draw [thick,dotted] (10.55,0.375) -- (10.80,0.375);

\draw [thick,dotted] (0.2,2) -- (0.45,2);
\draw [ultra thick] (0.55,1.5) -- (0.95,1.5);
\draw [ultra thick] (1.05,2.5) -- (1.45,2.5);
\draw [ultra thick] (1.55,1.75) -- (1.95,1.75);
\draw [ultra thick] (2.05,2.75) -- (2.45,2.75);
\draw [ultra thick] (2.55,1.75) -- (2.95,1.75);
\draw [ultra thick] (3.05,2.5) -- (3.45,2.5);
\draw [ultra thick] (3.55,1.5) -- (3.95,1.5);
\draw [ultra thick] (4.05,2.25) -- (4.45,2.25);
\draw [ultra thick] (4.55,1.5) -- (4.95,1.5);
\draw [ultra thick] (5.05,2.75) -- (5.45,2.75);
\draw [ultra thick] (5.55,1.5) -- (5.95,1.5);
\draw [ultra thick] (6.05,2.5) -- (6.45,2.5);
\draw [ultra thick] (6.55,1.75) -- (6.95,1.75);
\draw [ultra thick] (7.05,2.75) -- (7.45,2.75);
\draw [ultra thick] (7.55,1.5) -- (7.95,1.5);
\draw [ultra thick] (8.05,2.5) -- (8.45,2.5);
\draw [ultra thick] (8.55,1.75) -- (8.95,1.75);
\draw [ultra thick] (9.05,2.75) -- (9.45,2.75);
\draw [ultra thick] (9.55,2) -- (9.95,2);
\draw [ultra thick] (10.05,2.75) -- (10.45,2.75);
\draw [thick,dotted] (10.55,2) -- (10.80,2);

\end{tikzpicture}

\caption{
\label{figure: coexistence perconlating non percolating}
Coexistence of classes where percolation does and does not occur, for $d=1$, $\NumLevels \ge 3$. 
All configurations where a single site hosts one particle and all other sites are unoccupied, sit in the same class as $\eta_{(1)}$. 
There is percolation in this class, and the resonant dynamics restricted to it is in fact ballistic (restriction of $U_{res}$ is equivalent to the lattice Laplacian).
The state $\eta_{(2)}$ is such that neighboring sites always have a difference in occupation number larger than two. There is not a single resonant spot and $\eta_{(2)}$ is the only configuration in its class. 
}
\end{figure}
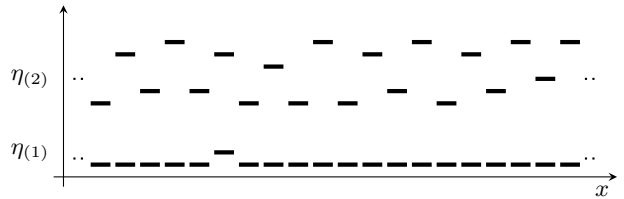

\section{Percolation of Resonances}\label{section: Percolation resonances Model I}

We investigate whether the resonances percolate in configuration space, for large volume $V$, that is, whether most configurations $\eta$ belong to a percolating class $\class$.
We first address this question in first order in  $\coupling$.    
Then, the answer is that, taking the Hamiltonian \eqref{original Hamiltonian} in $d=1$, 
there is no percolation, see Section \ref{sec: example without percolation}. 
Instead, taking $d\ge 2$, or even a strip of two lanes, or allowing for next-to-nearest neighbor hopping in $d=1$, there is percolation, see Section \ref{sec: example with percolation}. 
Then, in Section \ref{sec: nonperturbative}, we investigate non-perturbative effects and we argue that eventually there is percolation in all cases.

\subsection{Example without percolation in first order}\label{sec: example without percolation}
We first take the model to be precisely given by \eqref{original Hamiltonian} in $d=1$, so that the resonant Hamiltonian is given by, with $n_x=b^*_xb_x$, 
\begin{equation}\label{resonant Hamiltonian model I}
U_{res} 
\; = \;  
\sum_{x\in V} U_{res,x}
\; = \; 
\sum_{x\in V}  \big(  b_x^* \mathbbm{1}_{n_x=n_{x+1}} b_{x+1} +\text{h.c.}
\big).
\end{equation}
We prove that, for $\NumLevels$ not too small, most configurations are in a class $\class$ without percolation, 
More precisely,  we show in Appendix \ref{appendix: resonances percolation original model} that, if  the configuration $\eta$ satisfies
\begin{equation}\label{eq: condition non overtake}   
|\eta_x - \eta_{x-1}| \ge 3 \qquad  \text{and} \qquad |\eta_x - \eta_{x+1}| \ge 3,
\end{equation}
then the site $x$ if frozen in $\eta$, i.e.\@ for any $\eta' \in \class(\eta)$, it holds that $\eta'_x = \eta_x$. 
The proof is illustrated on figure \ref{figure: non-percolation for model I} in Appendix \ref{appendix: resonances percolation original model}: 
it is impossible to swap occupation numbers between two sites, if their difference is larger than one. 
We immediately see that for a typical configuration $\eta$, condition \eqref{eq: condition non overtake} is satisfied for a fraction of sites no less than  $1-C/\NumLevels$. 

For pedagogical
reasons to become clear in Section \ref{sec: nonperturbative}, we {also} introduce a small modification of our original $d=1$ model. 
Namely we add a two-boson hopping term so that now
\begin{equation*}
 U  \; = \;   \sum_{x\in V} \big(   b^*_{x} b_{x+1} {+}  (b^*_{x})^2 (b_{x+1})^2 + \text{h.c.}\big)
\end{equation*}
and the corresponding resonant Hamiltonian is  $U_{res}=\sum_{x\in V} U_{res,x}$ with 
 \beq \label{resonant Hamiltonian model I double}
 U_{res,x}
\; = \;   b_{x}^* \mathbbm{1}_{n_x=n_{x+1}} b_{x+1}  +    (b^*_{x})^2   \mathbbm{1}_{n_x=n_{x+1}}  (b_{x+1})^2 + \text{h.c.}
\eeq
In this case, an obvious modification of the proof {in Appendix \ref{appendix: resonances percolation original model}} applies and we see that, if 
\begin{equation}\label{eq: condition non overtake II}   |\eta_x - \eta_{x-1}| \ge 5 \qquad  \text{and} \qquad |\eta_x - \eta_{x+1}| \ge 5,  \end{equation}
holds, then $x$ is frozen.

\subsection{Examples with percolation in first order}\label{sec: example with percolation}
First, we consider again $d=1$ but we add a hopping term between next-to-nearest neighbors in the Hamiltonian \eqref{original Hamiltonian}.
This results in a resonant Hamiltonian of the form  $U_{res} 
\; = \;  
\sum_{x\in V} U_{res,x}$ with 
\begin{equation}\label{resonant Hamiltonian model I bis}
 U_{res,x}
\; = \; 
\sum_{x\in V}  \big( b_x^*  \mathbbm{1}_{n_x=n_{x+1}} b_{x+1} + b_x^* \mathbbm{1}_{n_x=n_{x+2}}   b_{x+2} + \text{h.c.}
\big).
\end{equation}
Second, we take the Hamiltonian \eqref{original Hamiltonian} on a two-lane strip:
\begin{equation*}
S = \{ (j,0),(j,1) : j\in I \subset \Z \}, 
\end{equation*}
giving rise to the resonant Hamiltonian 
\begin{equation}\label{resonant Hamiltonian model I ter}
U_{res} 
\; = \;  
\sum_{x\in S} U_{res,x}
\; = \; 
\frac{1}{3} \sum_{x\in S} \sum_{y \sim x}   \big(b_x^* \mathbbm{1}_{n_x=n_{y}} b_{y} +  \text{h.c.}\big).
\end{equation}
Larger strips or dimensions larger than one could be considered too. 
In all these cases, an overwhelming majority of configurations $\eta$ belongs to a class with percolation, as soon as  the volume $V$ is large enough (compared to $\NumLevels$). 
We postpone the proof of this claim to Appendix \ref{appendix: proof of percolation}, but we explain the idea here.
From the discussion of Section \ref{sec: many body localized phase}, we know that at low density, our system is ergodic (delocalized).  
Therefore, we expect that rare regions of low particle density will behave in an ergodic way as well, and that they could delocalize the system.  
We thus consider a subvolume $\Lambda \subset V$ with a local configuration satisfying $\sum_x \eta(x) = \rho \str \Lambda \str $ with $\rho < 1 \ll \NumLevels$. 
One of such configurations, that we call $\caF$, is depicted on figure \ref{figure: ergodic spot}, with $p\sim 1/\rho$. 
Note that  $|\Lambda | \ge p^{ \NumLevels +1}$, so that any occupation number appears in the spot, which is necessary for the mobility. 
This spot is shown to be able to travel across the system {if $p\geq 3$}, i.e.\ we can move $\caF$ to a translate of $\Lambda$,  as well as move particles from any place to any other one.   
The way the spot moves is strongly reminiscent of motion in kinetically constrained models \cite{KobAndersen}\cite{BiroliMezard}.  

It is worth pointing out that, since  the length of the ergodic spot is $|\Lambda | \sim C^{ \NumLevels }$, we need a volume $V \gg \NumLevels^{C | \Lambda|}$ for such a low-density spot to become typical. Therefore, percolation  is only established by the above argument for $V \geq  C^{C^{\NumLevels}}$.

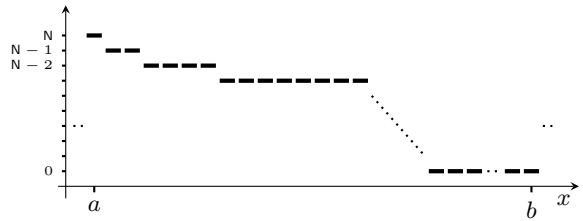
\begin{figure}[t]
\begin{tikzpicture}[xscale=0.5,yscale=0.8]

\draw [>=stealth,->] (-0.2,0.5) -- (13.5,0.5);
\draw [>=stealth,->] (0,0.3) -- (0,3.5);

\draw (13.1,0.5) node[below]{$x$} ;

\draw [thick] (-0.1,3) -- (0,3);
\draw [thick] (-0.1,2.75) -- (0,2.75);
\draw [thick] (-0.1,2.5) -- (0,2.5);
\draw [thick] (-0.1,2.25) -- (0,2.25);
\draw [thick] (-0.1,2) -- (0,2);
\draw [thick] (-0.1,1.75) -- (0,1.75);
\draw [thick] (-0.1,1.5) -- (0,1.5);
\draw [thick] (-0.1,1.25) -- (0,1.25);
\draw [thick] (-0.1,1) -- (0,1);
\draw [thick] (-0.1,0.75) -- (0,0.75);

\draw [thick] (0.75,0.4) -- (0.75,0.5);
\draw [thick] (12.25,0.4) -- (12.25,0.5);
\draw (0.75,0.4) node[below]{$a$} ;
\draw (12.25,0.4) node[below]{$b$} ;

\draw (-0.1,3) node[left]{\tiny$\NumLevels$};
\draw (-0.1,2.75) node[left]{\tiny$\NumLevels - 1$};
\draw (-0.1,2.5) node[left]{\tiny$\NumLevels - 2$};
\draw (-0.1,0.75) node[left]{\tiny$0$};

\draw [thick,dotted] (0.2,1.5) -- (0.45,1.5);
\draw [ultra thick] (0.55,3) -- (0.95,3);
\draw [ultra thick] (1.05,2.75) -- (1.45,2.75);
\draw [ultra thick] (1.55,2.75) -- (1.95,2.75);
\draw [ultra thick] (2.05,2.5) -- (2.45,2.5);
\draw [ultra thick] (2.55,2.5) -- (2.95,2.5);
\draw [ultra thick] (3.05,2.5) -- (3.45,2.5);
\draw [ultra thick] (3.55,2.5) -- (3.95,2.5);
\draw [ultra thick] (4.05,2.25) -- (4.45,2.25);
\draw [ultra thick] (4.55,2.25) -- (4.95,2.25);
\draw [ultra thick] (5.05,2.25) -- (5.45,2.25);
\draw [ultra thick] (5.55,2.25) -- (5.95,2.25);
\draw [ultra thick] (6.05,2.25) -- (6.45,2.25);
\draw [ultra thick] (6.55,2.25) -- (6.95,2.25);
\draw [ultra thick] (7.05,2.25) -- (7.45,2.25);
\draw [ultra thick] (7.55,2.25) -- (7.95,2.25);
\draw [thick,dotted] (8.05,2) -- (9.45,1);
\draw [ultra thick] (9.55,0.75) -- (9.95,0.75);
\draw [ultra thick] (10.05,0.75) -- (10.45,0.75);
\draw [ultra thick] (10.55,0.75) -- (10.95,0.75);
\draw [thick,dotted] (11.09,0.75) -- (11.42,0.75);
\draw [ultra thick] (11.55,0.75) -- (11.95,0.75);
\draw [ultra thick] (12.05,0.75) -- (12.45,0.75);
\draw [thick,dotted] (12.55,1.5) -- (12.80,1.5);

\end{tikzpicture}

\caption{
\label{figure: ergodic spot}
An ergodic spot $\mathcal F$ for the resonant Hamiltonian \eqref{resonant Hamiltonian model I bis}, delimited by the points $a$ and $b$. 
Let $p \ge 2$ ($p=2$ on the figure). 
The first site on the left has maximal occupation number $\NumLevels$, 
the next $p$ sites have occupation number $\NumLevels - 1$, 
the next $p^2$ sites have occupation number $\NumLevels - 2$, $\dots$, the last $p^{\NumLevels}$ sites are vacant. 
Therefore {the size of the spot $\mathcal F$ is} ${b-a} = (p^{\NumLevels +1} - 1) / (p-1)$. 
}
\end{figure}

\subsection{Nonperturbative argument for percolation}\label{sec: nonperturbative}

Up to now, we have investigated the role and mobility of resonances in first order in the hopping $\coupling$. 
In the cases where we found that resonances do not percolate, so in the examples of Section \ref{sec: example without percolation}, we can try to repeat the analysis at the second step of the scheme described in Section \ref{subsection: Basic picture RG}, as we show now. 
It is worth pointing out that this second step analysis does not coincide with a naive second order analysis, as non-perturbative effects are incorporated through the rotation $\tilde\Upsilon$; 
precisely these effects do allow us to establish in a robust way that percolation holds in great generality (similar considerations have been developed by \cite{HuseNandkishore}).

For the sake of simplicity, we consider the second model of Section \ref{sec: example without percolation}, i.e.\ with the $(b^*)^2b^2$-terms (in Appendix \ref{appendix: proof of non pert percolation} we argue that the same idea applies to the first model of Section \ref{sec: example without percolation}, or any model with short range interaction, for that matter).  After applying the transformation $\tilde \Omega$, introduced in Section \ref{subsection: Basic picture RG}, to the Hamiltonian $H$, we obtain the transformed Hamiltonian 
\begin{eqnarray} 
H' &=& \tilde\Omega^* H   \tilde \Omega =   H'_0 + \coupling^2 U'  \nonumber \\[1mm]
&=&  (E^{(0)} + \coupling U_{res}) + \tilde\Omega^* (\coupling U_{per})   \tilde \Omega   \label{eq: ham after first rotation}
\end{eqnarray}
We denote eigenstates of $H'_0$ by $\Psi,\Psi',\ldots$.   
Since $H'_0$ was found to be localized, the eigenstates can be written as $\Psi= \tilde\Upsilon\str \eta \rangle$  for some $\eta$, with $\tilde\Upsilon$ (see Section \ref{sec: resonances quenched})  identity in most places and non-local at rare resonant spots.  
We now implement the same percolation analysis as in Sections \ref{sec: example without percolation},\ref{sec: example with percolation}.  We declare a pair $(\Psi,\Psi')$ resonant if
 \beq \label{eq: condition for fusing}
\str \langle \Psi \str \coupling^2 U' \str   \Psi' \rangle \str  \; \gg \; \str \langle \Psi \str H_0' \str   \Psi \rangle - \langle \Psi' \str H_0' \str   \Psi' \rangle \str,
\eeq
in analogy to the condition \eqref{def: first order resonance} of the first step.  
We choose a $\Psi$ that has a resonant spot $S^*=[a,b]$ (discrete interval), with size roughly $\str b-a \str \geq C \big|  \log(\coupling  \NumLevels^{3/4}) \big| $  and particle density of order $1$}.  In the thermodynamic limit $V\to \infty$, an overwhelming majority of the eigenstates contain such a spot.
Assuming ergodicity in the resonant spot (more precisely, assuming ETH), we can find sequences $\Psi_1,\ldots, \Psi_n$ with $\Psi_1=\Psi$,  $(\Psi_i,\Psi_{i+1})$ resonant pairs, and $n \sim \NumLevels$, and such that $\Psi_n$ now contains {the resonant spot} shifted to $[a-1,b-1]$.
By increasing $S^*$ further, we can improve the inequality \eqref{eq: condition for fusing} and make the ratio of left-hand side to right hand side as large as desired, and we can exponentially {(in the size of $S^*$)} increase the number of choices for the sequence $\Psi_2,\ldots, \Psi_n$. We have gathered all details of this quite straightforward analysis in Appendix \ref{appendix: proof of non pert percolation}. 
In fact, we can simply summarize it by saying that a large ergodic spot can act as a thermal baths for the localized sites next to it.
Of course, by the same token, we can then connect $\Psi$ via resonant transitions to some $\Psi'$ having {the} ergodic spot in any desired place. {Likewise, we can also slightly change the number of particles in the spot, as long as the density is low enough to remain in the ergodic phase and the spot is large enough so that \eqref{eq: condition for fusing} holds. 
Therefore, the ergodic spot can transport particles from one place to another and $\Psi$ can eventually be connected resonantly to an overwhelming majority of the eigenstates.}
This suggests that ergodic bubbles can act as mobile carriers of  particles and energy and destroy the localization.

\subsection{From percolation to delocalization}
Finally, we come to the question whether percolation necessarily entails delocalization.  
Strictly speaking, all what we have argued is that the system does not manifestly break up in decoupled systems in perturbation theory, i.e.\ that the Hamiltonian acts on a truly connected graph of many-body states and the connections do not come with any small parameter. 

However, in principle it is still conceivable that the system is localized by interference effects \cite{HuseNandkishore}, 
in the same way that the adjacency matrix of a graph can have localized eigenstates even if the graph is connected (`quantum percolation') \cite{ShapirAharonyHarris,Veselic}.  
More concretely, one could think that the ergodic spot itself will be the entity that gets localized in a disordered background. 
However, one should realize that the ergodic spot is not like a passive particle moving in a fixed background, rather, as it moves, it can rearrange the background at will. 

In addition, one could fear that the ergodic spot will grow by absorbing bosons, until its density becomes so large that the spot is not longer ergodic (a variant on this objection is: the ergodic spot will split into smaller spots that are too small to be mobile, as we saw above that mobility requires a minimal size). 
It is certainly true that these effects will happen in a dynamical description of the system, but we do not see how they could avoid mobility of the spots on very long time scales. 
Indeed, by reversibility (detailed balance) of the Hamiltonian dynamics, it must be true that during the time evolution starting from equilibrium, any transition occurs equally often as its time-reversal, hence if (mobile spot $\rightarrow$ immobile spot) occurs, then also (immobile spot $\rightarrow$ mobile spot). 

Currently, we are investigating these issues further, also numerically \cite{De Roeck Huveneers Mueller Schiulaz}.  
Apart from the interest in translation invariant localization, this might shed a new light on the localization-delocalization transition in weakly disordered systems.

\section{Conclusion}
We have analyzed a model of interacting bosons on the lattice, introduced in Section \ref{section: First model}.  This model can be considered as a promising candidate for localization without quenched disorder. The basic reason for this is that, in a typical initial state at high energy density, the site-dependent boson numbers provide an effective random potential that could play the role of quenched disorder. We quantified this by exhibiting the fact that `resonant spots' (places where resonant transitions can take place) are sparse.  In that respect, the model is similar to models of strongly disordered spin chains, where MBL is believed to occur. 

We then addressed more precisely the question whether an MBL phase can be realized in our model.  Some obvious counterarguments were formulated in Section \ref{sec: many body localized phase}. They can be summarized by saying that not all eigenstates can be localized and that, even for those that would be localized, the localization deteriorates in rare regions with low energy density.  However, we argued that these objections, although they exclude `full MBL', are a priori still compatible with MBL. We did this by providing an abstract charcterization of the putative MBL phase.   

Then, in Section \ref{section: RG picture and Resonances}, we adapted to our model an iterative diagonalization scheme from \cite{Imbrie}, based on perturbation theory in the hopping. 
The result, stated in Section \ref{section: Percolation resonances Model I}, is that for some versions of our model (most notably nearest neighbor hopping in spatial dimension higher than $1$), resonances do consitute a mechanism for delocalization already in first order. 
More precisely, our analysis led to the conclusion that all localized eigenstates (present when the hopping is switched off) should hybridize with each other. 
However, this analysis rests on a detailed `percolation' analysis of the `resonant graph' that does not always (i.e.\ for all versions of our model) apply in first order. 
We then discussed these latter cases (for which a first order analysis predicts localization). 
Taking non-perturbative effects into account, we found a generic argument leading to the conclusion that localized eigenstates should hybridize, 
where `generic' means that the analysis is no longer dependent on the fine properties of the model anymore. 
In both cases, the crucial phenomenon is the presence of rare resonant spots that are shown to be mobile, i.e.\ they can travel through the system and rearrange the state. 
When based on non-perturbative effects, this conclusion relies on the ergodicity of the resonant spots as only input.  
Since the mobile resonant spots provide a mechanism for thermalization and transport, 
our work suggests a scenario for delocalization, and hence absence of an MBL phase, in translation invariant models like ours.

\appendix

\section{No percolation: $U_{res}$ given by \eqref{resonant Hamiltonian model I}}\label{appendix: resonances percolation original model}
We use the fact that the system is one-dimensional and that the interaction is strictly between nearest neighboors to deduce the following result. 
Consider two configurations $\eta$ and $\eta'$ that belong to the same class $\class$. That is, there is a finite sequence $\eta_{(1)}, \dots , \eta_{(n)}$ such that
\begin{equation*}
(\eta, \eta_{(1)}) \, ,\, (\eta_{(1)}, \eta_{(2)})\, , \, \dots \, , \, (\eta_{(n-1)}, \eta_{(n)})\, , \,  (\eta_{(n)}, \eta') 
\end{equation*}
are resonant pairs.
Then, for any $x \in V$,
\begin{equation}\label{lack of percolation Model I}
{ \big( |\eta_{x-1} - \eta_{x}| \ge 3 ,\; |\eta_x - \eta_{x+1}| \ge 3 \big)  \;\Rightarrow\; \eta'_x = \eta_x . }
\end{equation}

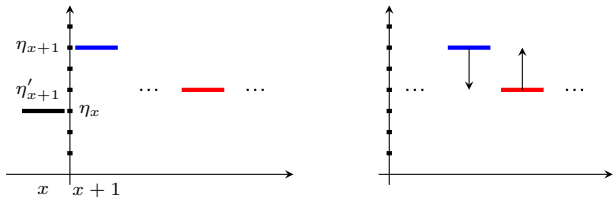
\begin{figure}[t]
\begin{tikzpicture}[scale=0.7]


\draw [>=stealth,->] (0.8,0) -- (6.2,0);
\draw [>=stealth,->] (2,-0.2) -- (2,3.2);

\draw (1.5,-0.05) node[below]{\scriptsize{$x$}} ;
\draw (2.5,0) node[below]{\scriptsize{$x+1$}} ;

\draw (2,2.4) node[left]{\scriptsize$\eta_{x+1}$} ;
\draw (2,1.6) node[left]{\scriptsize$\eta_{x+1}'$} ;
\draw (2,1.2) node[right]{\scriptsize$\eta_x$} ;

\draw [ultra thick] (1.95,0.4) -- (2.05,0.4);
\draw [ultra thick] (1.95,0.8) -- (2.05,0.8);
\draw [ultra thick] (1.95,1.2) -- (2.05,1.2);
\draw [ultra thick] (1.95,1.6) -- (2.05,1.6);
\draw [ultra thick] (1.95,2) -- (2.05,2);
\draw [ultra thick] (1.95,2.4) -- (2.05,2.4);
\draw [ultra thick] (1.95,2.8) -- (2.05,2.8);

\draw [ultra thick] (1.1,1.2) -- (1.9,1.2);
\draw [ultra thick,blue] (2.1,2.4) -- (2.9,2.4);
\draw [thick,dotted] (3.32,1.6) -- (3.69,1.6);
\draw [ultra thick,red] (4.1,1.6) -- (4.9,1.6);
\draw [thick,dotted] (5.32,1.6) -- (5.69,1.6);


\draw [>=stealth,->] (8,-0.2) -- (8,3.2);
\draw [>=stealth,->] (7.8,0) -- (12.2,0);

\draw [ultra thick] (7.95,0.4) -- (8.05,0.4);
\draw [ultra thick] (7.95,0.8) -- (8.05,0.8);
\draw [ultra thick] (7.95,1.2) -- (8.05,1.2);
\draw [ultra thick] (7.95,1.6) -- (8.05,1.6);
\draw [ultra thick] (7.95,2) -- (8.05,2);
\draw [ultra thick] (7.95,2.4) -- (8.05,2.4);
\draw [ultra thick] (7.95,2.8) -- (8.05,2.8);

\draw [thick,dotted] (8.32,1.6) -- (8.69,1.6);
\draw [ultra thick,blue] (9.1,2.4) -- (9.9,2.4);
\draw [ultra thick,red] (10.1,1.6) -- (10.9,1.6);
\draw [thick,dotted] (11.32,1.6) -- (11.69,1.6);

\draw[>=stealth,->] (9.5,2.4) -- (9.5,1.6);
\draw[>=stealth,->] (10.5,1.6) -- (10.5,2.4);

\end{tikzpicture}

\caption{
\label{figure: non-percolation for model I}
Validity of \eqref{lack of percolation Model I}: it is impossible to swap occupation numbers between two adjacent sites, if their difference is larger than one.
}
\end{figure}

{To show \eqref{lack of percolation Model I}, we refer to figure \ref{figure: non-percolation for model I}.
Let us assume $|\eta_{x-1} - \eta_{x}| \ge 3$ and  $|\eta_x - \eta_{x+1}| \ge 3$. 
Let us first remove the terms $U_{res,x-1}$ and $U_{res,x}$ from the Hamiltonian, so that the site $x$ is decoupled from the rest of the system. 
This implies $|\eta_{x+1} - \eta_{x+1}'| \le 1$ as, if  $\eta_{x+1} \ne \eta_{x+1}'$, 
the occupation number $\eta_{x+1}$ in blue on figure \ref{figure: non-percolation for model I} 
needs at some point to get swapped with the occupation number $\eta_{x+1}'$ in red figure \ref{figure: non-percolation for model I}. 
This can only be if $|\eta_{x+1} - \eta_{x+1}'| = 1$ (on figure \ref{figure: non-percolation for model I}, it is thus not possible). 
Similarly $|\eta_{x-1} - \eta_{x-1}'| \le 1$.
But then, we realize that the same conclusion could have been reached without removing the terms $U_{res,x-1}$ and $U_{res,x}$, implying that the site $x$ is frozen. }

\section{Percolation: $U_{res}$ given by \eqref{resonant Hamiltonian model I bis} and \eqref{resonant Hamiltonian model I ter}}\label{appendix: proof of percolation}
\begin{figure}[h!]

\begin{tikzpicture}[scale=0.55]

\draw (-0.8,0.15) node[left]{$(b)$} ;
\draw (-0.8,1.15) node[left]{$(a)$} ;


\draw[thick,dotted] (-0.6,1.15) -- (-0.25,1.15);

\draw[color=blue,fill=blue] (0,1) circle (0.1);
\draw[color=magenta,fill=magenta] (0.3,1) circle (0.1);
\draw[color=magenta,fill=magenta] (0.6,1) circle (0.1);
\draw[color=magenta,fill=magenta] (0.9,1) circle (0.1);
\draw[color=magenta,fill=magenta] (1.2,1) circle (0.1);
\draw[color=pink,fill=pink] (1.5,1) circle (0.1);
\draw[color=pink,fill=pink] (1.8,1) circle (0.1);
\draw[color=pink,fill=pink] (2.1,1) circle (0.1);
\draw[color=pink,fill=pink] (2.4,1) circle (0.1);
\draw[color=pink,fill=pink] (2.7,1) circle (0.1);
\draw[color=pink,fill=pink] (3,1) circle (0.1);
\draw[color=pink,fill=pink] (3.3,1) circle (0.1);
\draw[color=pink,fill=pink] (3.6,1) circle (0.1);
\draw[color=pink,fill=pink] (3.9,1) circle (0.1);
\draw[color=pink,fill=pink] (4.2,1) circle (0.1);
\draw[color=pink,fill=pink] (4.5,1) circle (0.1);
\draw[color=pink,fill=pink] (4.8,1) circle (0.1);
\draw[color=pink,fill=pink] (5.1,1) circle (0.1);
\draw[color=pink,fill=pink] (5.4,1) circle (0.1);
\draw[color=pink,fill=pink] (5.7,1) circle (0.1);
\draw[color=pink,fill=pink] (6,1) circle (0.1);

\draw[color=blue,fill=blue] (0,1.3) circle (0.1);
\draw[color=magenta,fill=magenta] (0.3,1.3) circle (0.1);
\draw[color=magenta,fill=magenta] (0.6,1.3) circle (0.1);
\draw[color=magenta,fill=magenta] (0.9,1.3) circle (0.1);
\draw[color=magenta,fill=magenta] (1.2,1.3) circle (0.1);
\draw[color=pink,fill=pink] (1.5,1.3) circle (0.1);
\draw[color=pink,fill=pink] (1.8,1.3) circle (0.1);
\draw[color=pink,fill=pink] (2.1,1.3) circle (0.1);
\draw[color=pink,fill=pink] (2.4,1.3) circle (0.1);
\draw[color=pink,fill=pink] (2.7,1.3) circle (0.1);
\draw[color=pink,fill=pink] (3,1.3) circle (0.1);
\draw[color=pink,fill=pink] (3.3,1.3) circle (0.1);
\draw[color=pink,fill=pink] (3.6,1.3) circle (0.1);
\draw[color=pink,fill=pink] (3.9,1.3) circle (0.1);
\draw[color=pink,fill=pink] (4.2,1.3) circle (0.1);
\draw[color=pink,fill=pink] (4.5,1.3) circle (0.1);
\draw[color=pink,fill=pink] (4.8,1.3) circle (0.1);
\draw[color=pink,fill=pink] (5.1,1.3) circle (0.1);
\draw[color=pink,fill=pink] (5.4,1.3) circle (0.1);
\draw[color=pink,fill=pink] (5.7,1.3) circle (0.1);
\draw[color=pink,fill=pink] (6,1.3) circle (0.1);

\draw[thick,dotted] (6.35,1.15) -- (6.7,1.15);

\draw[color=golden,fill=golden] (7,1) circle (0.1);
\draw[color=golden,fill=golden] (7.3,1) circle (0.1);
\draw[color=golden,fill=golden] (7.6,1) circle (0.1);
\draw[color=golden,fill=golden] (7.9,1) circle (0.1);
\draw[color=golden,fill=golden] (8.2,1) circle (0.1);
\draw[color=golden,fill=golden] (8.5,1) circle (0.1);
\draw[color=golden,fill=golden] (8.8,1) circle (0.1);
\draw[color=golden,fill=golden] (9.1,1) circle (0.1);
\draw[color=golden,fill=golden] (9.4,1) circle (0.1);
\draw[color=golden,fill=golden] (9.7,1) circle (0.1);

\draw[color=golden,fill=golden] (7,1.3) circle (0.1);
\draw[color=golden,fill=golden] (7.3,1.3) circle (0.1);
\draw[color=golden,fill=golden] (7.6,1.3) circle (0.1);
\draw[color=golden,fill=golden] (7.9,1.3) circle (0.1);
\draw[color=golden,fill=golden] (8.2,1.3) circle (0.1);
\draw[color=golden,fill=golden] (8.5,1.3) circle (0.1);
\draw[color=golden,fill=golden] (8.8,1.3) circle (0.1);
\draw[color=golden,fill=golden] (9.1,1.3) circle (0.1);
\draw[color=golden,fill=golden] (9.4,1.3) circle (0.1);
\draw[color=golden,fill=golden] (9.7,1.3) circle (0.1);

\draw[thick,dotted] (9.88,1.15) -- (10.12,1.15);

\draw[color=golden,fill=golden] (10.3,1) circle (0.1);
\draw[color=golden,fill=golden] (10.6,1) circle (0.1);
\draw[color=golden,fill=golden] (10.9,1) circle (0.1);
\draw[color=golden,fill=golden] (11.2,1) circle (0.1);
\draw[color=golden,fill=golden] (11.5,1) circle (0.1);
\draw[color=golden,fill=golden] (11.8,1) circle (0.1);
\draw[color=golden,fill=golden] (12.1,1) circle (0.1);

\draw[color=golden,fill=golden] (10.3,1.3) circle (0.1);
\draw[color=golden,fill=golden] (10.6,1.3) circle (0.1);
\draw[color=golden,fill=golden] (10.9,1.3) circle (0.1);
\draw[color=golden,fill=golden] (11.2,1.3) circle (0.1);
\draw[color=golden,fill=golden] (11.5,1.3) circle (0.1);
\draw[color=golden,fill=golden] (11.8,1.3) circle (0.1);
\draw[color=golden,fill=golden] (12.1,1.3) circle (0.1);

\draw[thick,dotted] (12.4,1.15) -- (12.76,1.15);


\draw[thick,dotted] (-0.6,0.15) -- (-0.25,0.15);

\draw[color=pink,fill=pink] (0,0) circle (0.1);
\draw[color=blue,fill=blue] (0.3,0) circle (0.1);
\draw[color=pink,fill=pink] (0.6,0) circle (0.1);
\draw[color=pink,fill=pink] (0.9,0) circle (0.1);
\draw[color=pink,fill=pink] (1.2,0) circle (0.1);
\draw[color=pink,fill=pink] (1.5,0) circle (0.1);
\draw[color=pink,fill=pink] (1.8,0) circle (0.1);
\draw[color=pink,fill=pink] (2.1,0) circle (0.1);
\draw[color=magenta,fill=magenta] (2.4,0) circle (0.1);
\draw[color=pink,fill=pink] (2.7,0) circle (0.1);
\draw[color=pink,fill=pink] (3,0) circle (0.1);
\draw[color=pink,fill=pink] (3.3,0) circle (0.1);
\draw[color=magenta,fill=magenta] (3.6,0) circle (0.1);
\draw[color=pink,fill=pink] (3.9,0) circle (0.1);
\draw[color=pink,fill=pink] (4.2,0) circle (0.1);
\draw[color=pink,fill=pink] (4.5,0) circle (0.1);
\draw[color=magenta,fill=magenta] (4.8,0) circle (0.1);
\draw[color=pink,fill=pink] (5.1,0) circle (0.1);
\draw[color=pink,fill=pink] (5.4,0) circle (0.1);
\draw[color=pink,fill=pink] (5.7,0) circle (0.1);
\draw[color=magenta,fill=magenta] (6,0) circle (0.1);

\draw[color=pink,fill=pink] (0,0.3) circle (0.1);
\draw[color=pink,fill=pink] (0.3,0.3) circle (0.1);
\draw[color=pink,fill=pink] (0.6,0.3) circle (0.1);
\draw[color=blue,fill=blue] (0.9,0.3) circle (0.1);
\draw[color=pink,fill=pink] (1.2,0.3) circle (0.1);
\draw[color=pink,fill=pink] (1.5,0.3) circle (0.1);
\draw[color=magenta,fill=magenta] (1.8,0.3) circle (0.1);
\draw[color=pink,fill=pink] (2.1,0.3) circle (0.1);
\draw[color=pink,fill=pink] (2.4,0.3) circle (0.1);
\draw[color=pink,fill=pink] (2.7,0.3) circle (0.1);
\draw[color=magenta,fill=magenta] (3,0.3) circle (0.1);
\draw[color=pink,fill=pink] (3.3,0.3) circle (0.1);
\draw[color=pink,fill=pink] (3.6,0.3) circle (0.1);
\draw[color=pink,fill=pink] (3.9,0.3) circle (0.1);
\draw[color=magenta,fill=magenta] (4.2,0.3) circle (0.1);
\draw[color=pink,fill=pink] (4.5,0.3) circle (0.1);
\draw[color=pink,fill=pink] (4.8,0.3) circle (0.1);
\draw[color=pink,fill=pink] (5.1,0.3) circle (0.1);
\draw[color=magenta,fill=magenta] (5.4,0.3) circle (0.1);
\draw[color=pink,fill=pink] (5.7,0.3) circle (0.1);
\draw[color=pink,fill=pink] (6,0.3) circle (0.1);

\draw[thick,dotted] (6.35,0.15) -- (6.7,0.15);

\draw[color=golden,fill=golden] (7,0) circle (0.1);
\draw[color=golden,fill=golden] (7.3,0) circle (0.1);
\draw[color=golden,fill=golden] (7.6,0) circle (0.1);
\draw[color=golden,fill=golden] (7.9,0) circle (0.1);
\draw[color=golden,fill=golden] (8.2,0) circle (0.1);
\draw[color=golden,fill=golden] (8.5,0) circle (0.1);
\draw[color=golden,fill=golden] (8.8,0) circle (0.1);
\draw[color=golden,fill=golden] (9.1,0) circle (0.1);
\draw[color=golden,fill=golden] (9.4,0) circle (0.1);
\draw[color=golden,fill=golden] (9.7,0) circle (0.1);

\draw[color=golden,fill=golden] (7,0.3) circle (0.1);
\draw[color=golden,fill=golden] (7.3,0.3) circle (0.1);
\draw[color=golden,fill=golden] (7.6,0.3) circle (0.1);
\draw[color=golden,fill=golden] (7.9,0.3) circle (0.1);
\draw[color=golden,fill=golden] (8.2,0.3) circle (0.1);
\draw[color=golden,fill=golden] (8.5,0.3) circle (0.1);
\draw[color=golden,fill=golden] (8.8,0.3) circle (0.1);
\draw[color=golden,fill=golden] (9.1,0.3) circle (0.1);
\draw[color=golden,fill=golden] (9.4,0.3) circle (0.1);
\draw[color=golden,fill=golden] (9.7,0.3) circle (0.1);

\draw[thick,dotted] (9.88,0.15) -- (10.12,0.15);

\draw[color=golden,fill=golden] (10.3,0) circle (0.1);
\draw[color=golden,fill=golden] (10.6,0) circle (0.1);
\draw[color=golden,fill=golden] (10.9,0) circle (0.1);
\draw[color=golden,fill=golden] (11.2,0) circle (0.1);
\draw[color=golden,fill=golden] (11.5,0) circle (0.1);
\draw[color=golden,fill=golden] (11.8,0) circle (0.1);
\draw[color=golden,fill=golden] (12.1,0) circle (0.1);

\draw[color=golden,fill=golden] (10.3,0.3) circle (0.1);
\draw[color=golden,fill=golden] (10.6,0.3) circle (0.1);
\draw[color=golden,fill=golden] (10.9,0.3) circle (0.1);
\draw[color=golden,fill=golden] (11.2,0.3) circle (0.1);
\draw[color=golden,fill=golden] (11.5,0.3) circle (0.1);
\draw[color=golden,fill=golden] (11.8,0.3) circle (0.1);
\draw[color=golden,fill=golden] (12.1,0.3) circle (0.1);

\draw[thick,dotted] (12.4,0.15) -- (12.76,0.15);

\end{tikzpicture}

\caption{
\label{figure: ergodic spot model I ter}
ergodic spot for the resonant Hamiltonian \eqref{resonant Hamiltonian model I ter}. 
$(a)$ 
Sites in blue have occupation number $\NumLevels$, sites in magenta have occupation number $\NumLevels - 1$, $\dots$, sites in yellow are vacant. 
There are $p=4$ times more sites occupied by $k$ particles than sites occupied by $k+1$ particles. 
As a consequence, sites with $k+1$ particles can be diluted among sites with $k$ particles.
This spot can play for Hamiltonian \eqref{resonant Hamiltonian model I ter}, 
the role played by the spot depicted on figure \ref{figure: ergodic spot} for Hamiltonian \eqref{resonant Hamiltonian model I bis}.
$(b)$ Example of dilution of the sites with $\NumLevels$ and $\NumLevels - 1$ partilcles among sites with $\NumLevels - 2$ particles. 
}
\end{figure}
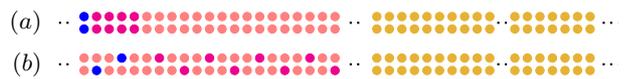

For the model with resonant Hamiltonian given by \eqref{resonant Hamiltonian model I bis},
we show that if two configurations $\eta,\eta'$ satisfy 
\begin{enumerate}
\item   
$\str \{ x \in V:  \eta_x=n \} \str  =   \str \{ x\in V:  \eta'_x=n \} \str $ for all $n=0,\ldots, \NumLevels$. 
\item  
They contain an `ergodic spot' $\mathcal F$ with $p=3$, as depicted on figure \ref{figure: ergodic spot}.
\end{enumerate}
then $\eta, \eta'$ belong to the same class $\class$. 
For large enough $V$, a typical configuration will contain 
a ergodic spot somewhere, so that most of the configurations belong to classes with percolation.  
The first condition above is merely a manifestation of the obvious `polynomial' constraint mentioned already in Section \ref{sec: Resonances: translation invariant systems}. 
For the model \eqref{resonant Hamiltonian model I ter}, the statement is the same except that the ergodic spot $\caF$ is now the one from figure \ref{figure: ergodic spot model I ter}.

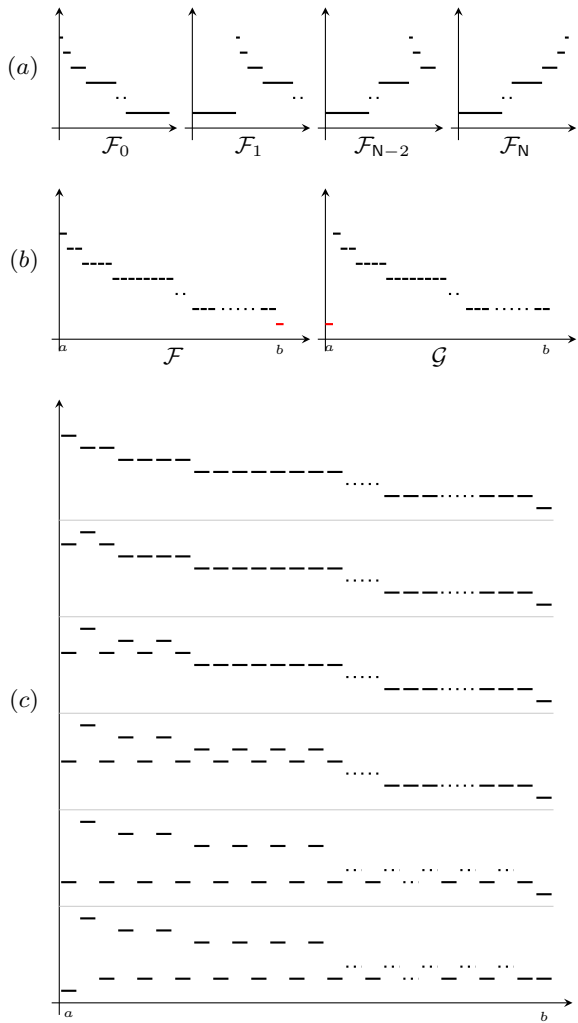
\begin{figure}[h!]


\begin{tikzpicture}[xscale=0.5,yscale=0.8]


\draw (-0.3,15.5) node[left]{$(a)$} ;

\draw [>=stealth,->] (-0.2,14.5) -- (3.1,14.5);
\draw [>=stealth,->] (0,14.3) -- (0,16.5);
\draw (1.5,14.5) node[below]{$\mathcal F_0$} ;

\draw [>=stealth,->] (3.3,14.5) -- (6.6,14.5);
\draw [>=stealth,->] (3.5,14.3) -- (3.5,16.5);
\draw (5,14.5) node[below]{$\mathcal F_1$} ;

\draw [>=stealth,->] (6.8,14.5) -- (10.1,14.5);
\draw [>=stealth,->] (7,14.3) -- (7,16.5);
\draw (8.5,14.5) node[below]{$\mathcal F_{\NumLevels - 2}$} ;

\draw [>=stealth,->] (10.3,14.5) -- (13.6,14.5);
\draw [>=stealth,->] (10.5,14.3) -- (10.5,16.5);
\draw (12,14.5) node[below]{$\mathcal F_{\NumLevels}$} ;

\draw[thick] (0,16) -- (0.1,16);
\draw[thick] (0.1,15.75) -- (0.3,15.75);
\draw[thick] (0.3,15.5) -- (0.7,15.5);
\draw[thick] (0.7,15.25) -- (1.5,15.25);
\draw[thick,dotted] (1.5,15) -- (1.75,15);
\draw[thick] (1.75,14.75) -- (2.9,14.75);

\draw[thick] (3.5,14.75) -- (4.65,14.75);
\draw[thick] (4.65,16) -- (4.75,16);
\draw[thick] (4.75,15.75) -- (4.95,15.75);
\draw[thick] (4.95,15.5) -- (5.35,15.5);
\draw[thick] (5.35,15.25) -- (6.15,15.25);
\draw[thick,dotted] (6.15,15) -- (6.4,15);

\draw[thick] (7,14.75) -- (8.15,14.75);
\draw[thick,dotted] (8.15,15) -- (8.4,15);
\draw[thick] (8.4,15.25) -- (9.2,15.25);
\draw[thick] (9.2,16) -- (9.3,16);
\draw[thick] (9.3,15.75) -- (9.5,15.75);
\draw[thick] (9.5,15.5) -- (9.9,15.5);

\draw[thick] (10.5,14.75) -- (11.65,14.75);
\draw[thick,dotted] (11.65,15) -- (11.9,15);
\draw[thick] (11.9,15.25) -- (12.7,15.25);
\draw[thick] (12.7,15.5) -- (13.1,15.5);
\draw[thick] (13.1,15.75) -- (13.3,15.75);
\draw[thick] (13.3,16) -- (13.4,16);


\draw [>=stealth,->] (-0.2,11) -- (6.6,11);
\draw [>=stealth,->] (0,10.8) -- (0,13.5);
\draw (-0.3,12.3) node[left]{$(b)$} ;

\draw (3,11) node[below]{$\mathcal F$} ;
\draw (0.1,11.05) node[below]{\tiny$a$} ;
\draw (5.8,11.05) node[below]{\tiny$b$} ;

\draw[thick] (0,12.75) -- (0.2,12.75);

\draw[thick] (0.2,12.5) -- (0.38,12.5) ; 
\draw[thick] (0.42,12.5)-- (0.6,12.5);

\draw[thick] (0.6,12.25) -- (0.78,12.25);
\draw[thick] (0.82,12.25) -- (0.98,12.25);
\draw[thick] (1.02,12.25) -- (1.18,12.25);
\draw[thick] (1.22,12.25) -- (1.38,12.25);

\draw[thick] (1.4,12) -- (1.58,12);
\draw[thick] (1.62,12) -- (1.78,12);
\draw[thick] (1.82,12) -- (1.98,12);
\draw[thick] (2.02,12) -- (2.18,12);
\draw[thick] (2.22,12) -- (2.38,12);
\draw[thick] (2.42,12) -- (2.58,12);
\draw[thick] (2.62,12) -- (2.78,12);
\draw[thick] (2.82,12) -- (3,12);

\draw[thick,dotted] (3.06,11.75) -- (3.43,11.75);

\draw[thick] (3.5,11.5) -- (3.68,11.5);
\draw[thick] (3.72,11.5) -- (3.88,11.5);
\draw[thick] (3.92,11.5) -- (4.1,11.5);
\draw[thick,dotted] (4.28,11.5) -- (5.15,11.5);
\draw[thick] (5.3,11.5) -- (5.48,11.5);
\draw[thick] (5.52,11.5) -- (5.7,11.5);

\draw[thick,red] (5.7,11.25) -- (5.9,11.25);

\draw [>=stealth,->] (6.8,11) -- (13.6,11);
\draw [>=stealth,->] (7,10.8) -- (7,13.5);

\draw (10,11) node[below]{$\mathcal G$} ;
\draw (7.1,11.05) node[below]{\tiny$a$} ;
\draw (12.8,11.05) node[below]{\tiny$b$} ;

\draw[thick,red] (7,11.25) -- (7.2,11.25);

\draw[thick] (7.2,12.75) -- (7.4,12.75);

\draw[thick] (7.4,12.5) -- (7.58,12.5) ; 
\draw[thick] (7.62,12.5)-- (7.8,12.5);

\draw[thick] (7.8,12.25) -- (7.98,12.25);
\draw[thick] (8.02,12.25) -- (8.18,12.25);
\draw[thick] (8.22,12.25) -- (8.38,12.25);
\draw[thick] (8.42,12.25) -- (8.6,12.25);

\draw[thick] (8.6,12) -- (8.78,12);
\draw[thick] (8.82,12) -- (8.98,12);
\draw[thick] (9.02,12) -- (9.18,12);
\draw[thick] (9.22,12) -- (9.38,12);
\draw[thick] (9.42,12) -- (9.58,12);
\draw[thick] (9.62,12) -- (9.78,12);
\draw[thick] (9.82,12) -- (9.98,12);
\draw[thick] (10.02,12) -- (10.18,12);

\draw[thick,dotted] (10.25,11.75) -- (10.58,11.75);

\draw[thick] (10.7,11.5) -- (10.88,11.5);
\draw[thick] (10.92,11.5) -- (11.08,11.5);
\draw[thick] (11.12,11.5) -- (11.3,11.5);
\draw[thick,dotted] (11.48,11.5) -- (12.35,11.5);
\draw[thick] (12.5,11.5) -- (12.68,11.5);
\draw[thick] (12.72,11.5) -- (12.9,11.5);

\draw [>=stealth,->] (-0.2,0) -- (13.5,0);
\draw [>=stealth,->] (0,-0.2) -- (0,10);
\draw (-0.3,5) node[left]{$(c)$} ;
\draw (0.25,0) node[below]{\tiny$a$} ;
\draw (12.75,0) node[below]{\tiny$b$} ;

\draw[thick] (0.05,9.4) -- (0.45,9.4);

\draw[thick] (0.55,9.2) -- (0.95,9.2);
\draw[thick] (1.05,9.2) -- (1.45,9.2);

\draw[thick] (1.55,9) -- (1.95,9);
\draw[thick] (2.05,9) -- (2.45,9);
\draw[thick] (2.55,9) -- (2.95,9);
\draw[thick] (3.05,9) -- (3.45,9);

\draw[thick] (3.55,8.8) -- (3.95,8.8);
\draw[thick] (4.05,8.8) -- (4.45,8.8);
\draw[thick] (4.55,8.8) -- (4.95,8.8);
\draw[thick] (5.05,8.8) -- (5.45,8.8);
\draw[thick] (5.55,8.8) -- (5.95,8.8);
\draw[thick] (6.05,8.8) -- (6.45,8.8);
\draw[thick] (6.55,8.8) -- (6.95,8.8);
\draw[thick] (7.05,8.8) -- (7.45,8.8);

\draw[thick,dotted] (7.55,8.6) -- (8.45,8.6);

\draw[thick] (8.55,8.4) -- (8.95,8.4);
\draw[thick] (9.05,8.4) -- (9.45,8.4);
\draw[thick] (9.55,8.4) -- (9.95,8.4);
\draw[thick,dotted] (10.05,8.4) -- (10.95,8.4);
\draw[thick] (11.05,8.4) -- (11.45,8.4);
\draw[thick] (11.55,8.4) -- (11.95,8.4);
\draw[thick] (12.05,8.4) -- (12.45,8.4);

\draw[thick] (12.55,8.2) -- (12.95,8.2);

\draw[gray!40] (0,8) -- (13,8);

\draw[thick] (0.05,7.6) -- (0.45,7.6);
\draw[thick] (0.55,7.8) -- (0.95,7.8);
\draw[thick] (1.05,7.6) -- (1.45,7.6);

\draw[thick] (1.55,7.4) -- (1.95,7.4);
\draw[thick] (2.05,7.4) -- (2.45,7.4);
\draw[thick] (2.55,7.4) -- (2.95,7.4);
\draw[thick] (3.05,7.4) -- (3.45,7.4);

\draw[thick] (3.55,7.2) -- (3.95,7.2);
\draw[thick] (4.05,7.2) -- (4.45,7.2);
\draw[thick] (4.55,7.2) -- (4.95,7.2);
\draw[thick] (5.05,7.2) -- (5.45,7.2);
\draw[thick] (5.55,7.2) -- (5.95,7.2);
\draw[thick] (6.05,7.2) -- (6.45,7.2);
\draw[thick] (6.55,7.2) -- (6.95,7.2);
\draw[thick] (7.05,7.2) -- (7.45,7.2);

\draw[thick,dotted] (7.55,7) -- (8.45,7);

\draw[thick] (8.55,6.8) -- (8.95,6.8);
\draw[thick] (9.05,6.8) -- (9.45,6.8);
\draw[thick] (9.55,6.8) -- (9.95,6.8);
\draw[thick,dotted] (10.05,6.8) -- (10.95,6.8);
\draw[thick] (11.05,6.8) -- (11.45,6.8);
\draw[thick] (11.55,6.8) -- (11.95,6.8);
\draw[thick] (12.05,6.8) -- (12.45,6.8);

\draw[thick] (12.55,6.6) -- (12.95,6.6);

\draw[gray!40] (0,6.4) -- (13,6.4);


\draw[thick] (0.05,5.8) -- (0.45,5.8);
\draw[thick] (0.55,6.2) -- (0.95,6.2);
\draw[thick] (1.05,5.8) -- (1.45,5.8);

\draw[thick] (1.55,6) -- (1.95,6);
\draw[thick] (2.05,5.8) -- (2.45,5.8);
\draw[thick] (2.55,6) -- (2.95,6);
\draw[thick] (3.05,5.8) -- (3.45,5.8);

\draw[thick] (3.55,5.6) -- (3.95,5.6);
\draw[thick] (4.05,5.6) -- (4.45,5.6);
\draw[thick] (4.55,5.6) -- (4.95,5.6);
\draw[thick] (5.05,5.6) -- (5.45,5.6);
\draw[thick] (5.55,5.6) -- (5.95,5.6);
\draw[thick] (6.05,5.6) -- (6.45,5.6);
\draw[thick] (6.55,5.6) -- (6.95,5.6);
\draw[thick] (7.05,5.6) -- (7.45,5.6);

\draw[thick,dotted] (7.55,5.4) -- (8.45,5.4);

\draw[thick] (8.55,5.2) -- (8.95,5.2);
\draw[thick] (9.05,5.2) -- (9.45,5.2);
\draw[thick] (9.55,5.2) -- (9.95,5.2);
\draw[thick,dotted] (10.05,5.2) -- (10.95,5.2);
\draw[thick] (11.05,5.2) -- (11.45,5.2);
\draw[thick] (11.55,5.2) -- (11.95,5.2);
\draw[thick] (12.05,5.2) -- (12.45,5.2);

\draw[thick] (12.55,5) -- (12.95,5);

\draw[gray!40] (0,4.8) -- (13,4.8);


\draw[thick] (0.05,4) -- (0.45,4);
\draw[thick] (0.55,4.6) -- (0.95,4.6);
\draw[thick] (1.05,4) -- (1.45,4);

\draw[thick] (1.55,4.4) -- (1.95,4.4);
\draw[thick] (2.05,4) -- (2.45,4);
\draw[thick] (2.55,4.4) -- (2.95,4.4);
\draw[thick] (3.05,4) -- (3.45,4);

\draw[thick] (3.55,4.2) -- (3.95,4.2);
\draw[thick] (4.05,4) -- (4.45,4);
\draw[thick] (4.55,4.2) -- (4.95,4.2);
\draw[thick] (5.05,4) -- (5.45,4);
\draw[thick] (5.55,4.2) -- (5.95,4.2);
\draw[thick] (6.05,4) -- (6.45,4);
\draw[thick] (6.55,4.2) -- (6.95,4.2);
\draw[thick] (7.05,4) -- (7.45,4);

\draw[thick,dotted] (7.55,3.8) -- (8.45,3.8);

\draw[thick] (8.55,3.6) -- (8.95,3.6);
\draw[thick] (9.05,3.6) -- (9.45,3.6);
\draw[thick] (9.55,3.6) -- (9.95,3.6);
\draw[thick,dotted] (10.05,3.6) -- (10.95,3.6);
\draw[thick] (11.05,3.6) -- (11.45,3.6);
\draw[thick] (11.55,3.6) -- (11.95,3.6);
\draw[thick] (12.05,3.6) -- (12.45,3.6);

\draw[thick] (12.55,3.4) -- (12.95,3.4);

\draw[gray!40] (0,3.2) -- (13,3.2);


\draw[thick] (0.05,2) -- (0.45,2);
\draw[thick] (0.55,3) -- (0.95,3);
\draw[thick] (1.05,2) -- (1.45,2);

\draw[thick] (1.55,2.8) -- (1.95,2.8);
\draw[thick] (2.05,2) -- (2.45,2);
\draw[thick] (2.55,2.8) -- (2.95,2.8);
\draw[thick] (3.05,2) -- (3.45,2);

\draw[thick] (3.55,2.6) -- (3.95,2.6);
\draw[thick] (4.05,2) -- (4.45,2);
\draw[thick] (4.55,2.6) -- (4.95,2.6);
\draw[thick] (5.05,2) -- (5.45,2);
\draw[thick] (5.55,2.6) -- (5.95,2.6);
\draw[thick] (6.05,2) -- (6.45,2);
\draw[thick] (6.55,2.6) -- (6.95,2.6);
\draw[thick] (7.05,2) -- (7.45,2);

\draw[thick,dotted] (7.55,2.2) -- (7.95,2.2);
\draw[thick] (8.05,2) -- (8.45,2);
\draw[thick,dotted] (8.55,2.2) -- (8.95,2.2);
\draw[thick,dotted] (9.05,2) -- (9.45,2);
\draw[thick,dotted] (9.55,2.2) -- (9.95,2.2);
\draw[thick] (10.05,2) -- (10.45,2);
\draw[thick,dotted] (10.55,2.2) -- (10.95,2.2);
\draw[thick] (11.05,2) -- (11.45,2);
\draw[thick,dotted] (11.55,2.2) -- (11.95,2.2);
\draw[thick] (12.05,2) -- (12.45,2);

\draw[thick] (12.55,1.8) -- (12.95,1.8);

\draw[gray!40] (0,1.6) -- (13,1.6);


\draw[thick] (0.05,0.2) -- (0.45,0.2);
\draw[thick] (0.55,1.4) -- (0.95,1.4);
\draw[thick] (1.05,0.4) -- (1.45,0.4);

\draw[thick] (1.55,1.2) -- (1.95,1.2);
\draw[thick] (2.05,0.4) -- (2.45,0.4);
\draw[thick] (2.55,1.2) -- (2.95,1.2);
\draw[thick] (3.05,0.4) -- (3.45,0.4);

\draw[thick] (3.55,1) -- (3.95,1);
\draw[thick] (4.05,0.4) -- (4.45,0.4);
\draw[thick] (4.55,1) -- (4.95,1);
\draw[thick] (5.05,0.4) -- (5.45,0.4);
\draw[thick] (5.55,1) -- (5.95,1);
\draw[thick] (6.05,0.4) -- (6.45,0.4);
\draw[thick] (6.55,1) -- (6.95,1);
\draw[thick] (7.05,0.4) -- (7.45,0.4);

\draw[thick,dotted] (7.55,0.6) -- (7.95,0.6);
\draw[thick] (8.05,0.4) -- (8.45,0.4);
\draw[thick,dotted] (8.55,0.6) -- (8.95,0.6);
\draw[thick,dotted] (9.05,0.4) -- (9.45,0.4);
\draw[thick,dotted] (9.55,0.6) -- (9.95,0.6);
\draw[thick] (10.05,0.4) -- (10.45,0.4);
\draw[thick,dotted] (10.55,0.6) -- (10.95,0.6);
\draw[thick] (11.05,0.4) -- (11.45,0.4);
\draw[thick,dotted] (11.55,0.6) -- (11.95,0.6);
\draw[thick] (12.05,0.4) -- (12.45,0.4);

\draw[thick] (12.55,0.4) -- (12.95,0.4);

\end{tikzpicture}

\caption{
\label{figure: flip ergodic spot}
The spots $\mathcal F=\mathcal F_0,\mathcal F_1,\dots , \mathcal F_\NumLevels$ ($p=2$).
All what is depicted on the figure takes place inside the volume delimited by $a$ and $b$. 
Panel $(a)$. We aim to show that the spots $\mathcal F_0, \dots , \mathcal F_\NumLevels$ are connected.  
Panel $(b)$. It is enough to establish that $\mathcal F$ and $\mathcal G$ are connected, as the procedure can then be iterated.
Panel $(c)$. Half of the way from $\mathcal F$ to $\mathcal G$: the vacancy is transfered from the most right atom to most left site. 
Once this is accomplished, it is realized that all moves can be undone, wihle letting the leftmost site vacant. 
So it is actually seen how to move from $\mathcal F$ to $\mathcal G$, hence from $\mathcal F$ to $\mathcal F_1 , \dots , \mathcal F_\NumLevels$.  
The case $p> 2$ is analogous.
}
\end{figure}

Let us first consider \eqref{resonant Hamiltonian model I bis}. Let us consider the ergodic spot $\mathcal F$ shown on figure \ref{figure: ergodic spot}, for $p=2,3$.
The spot $\mathcal F$ is delimited in space by two sites $a$ and $b$, that we take fixed for the moment. 
On figure \ref{figure: flip ergodic spot}, 
it is shown that $\mathcal F$ is connected via resonant transitions to several other configurations $\mathcal F_1, \dots , \mathcal F_{\NumLevels}$, 
living in the same volume delimited by $a$ and $b$. 
While this alone does not entail percolation, simple generalizations of our construction will do.

Indeed, let us first see that $\mathcal F$ can travel through the chain if $p\ge 3$, which means that the two configurations (identical for all $x \le a-1$ and all $x \ge b+2$)
\begin{multline}\label{ergodic spot mechanism 1}
(\dots, \eta_{a-1},\mathcal F,\eta_{b+1},\eta_{b+2},\dots )
\qquad \text{and}\\[1mm]
(\dots, \eta_{a-1},\eta_{b+1},\mathcal F,\eta_{b+2},\dots )
\end{multline}
are connected via resonant transitions. 
For this, we first transform $\mathcal F$ into $\mathcal F_{\eta_{b+1}}$ as depticted on the figure \ref{figure: flip ergodic spot} $(a)$.
At this point, we absorb the site $b+1$ into the spot, and then undo all the previous steps, which is possible for $p\ge 3$. 
Doing so, we come back to $\tilde{\mathcal F}$ instead of $\mathcal F$, 
a spot that looks like $\mathcal F$, except that it lives in the interval $[a,b+1]$ instead of $[a,b]$, and that it contains one more site with occupation number $\eta_{b+1}$.  
We now need to evacuate this occupation number to the left side. 
For this we do the successive transformations represented in figure \ref{figure: flip ergodic spot} $(c)$, up to the moment that the site $a$ has occupation number $\eta_{b+1}$. 
We let then site $a$ with occupation number $\eta_{b+1}$, and we undo the other changes. 
We end up with the state $(\dots, \eta_{a-1},\eta_{b+1},\mathcal F,\eta_{b+2},\dots )$.

Second, $\mathcal F$ can be used to swap the occupation number of two near sites if $p\ge 3$, whatever these numbers are: the two states
\begin{equation}\label{ergodic spot mechanism 2}
(\dots, \eta_{a-1},\mathcal F, \eta_{b+1}, \dots )
,\;
(\dots, \eta_{b+1},\mathcal F, \eta_{a-1}, \dots )
\end{equation}
are connected via resonances, for any value of $\eta_{a-1}$ and $\eta_{b+1}$.  
To see this, we apply just a variant of the shceme leading to \eqref{ergodic spot mechanism 1}.  
We make the steps illustrated on figure \ref{figure: flip ergodic spot} $(a)$ and $(c)$ to absorb the two sites $a-1$ and $b+1$ into the spot, 
then undo the steps to get a spot $\tilde{\mathcal F}$ living on the interval $[a-1,b+1]$ and contaning one extra site with occupation number $\eta_{a-1}$ 
and one extra site with occupation number $\eta_{b+1}$. 
This procedure is then repeated, this time to evacuate the occupation number $\eta_{a-1}$ on the right and the occupation number $\eta_{b+1}$ on the left. 

It is finally seen that, combining mechanisms \eqref{ergodic spot mechanism 1} and \eqref{ergodic spot mechanism 2}, 
it becomes possible to permute the occupation number of any site with the occupation number any other one. 
We so arrive to the desired conclusion for the Hamiltonian given by \eqref{resonant Hamiltonian model I bis}. 

To deal with the Hamiltonian given by \eqref{resonant Hamiltonian model I ter}, a corresponding ergodic spot is constructed in figure \ref{figure: ergodic spot model I ter}. 
An inspection of this figure shows that this spot can play the same role as the ergodic spot used for \eqref{resonant Hamiltonian model I bis}, and the proof is concluded in an analogous way.

\section{Nonperturbative percolation: Proof of claims in Section \ref{sec: nonperturbative} }\label{appendix: proof of non pert percolation}

%
%

\paragraph{Outcome of the first step of the renormalization scheme.}
Let us first display $ H'_0 + \coupling^2 U' =  (E^{(0)} + \coupling U_{res}) + \tilde\Omega^* (\coupling U_{per})   \tilde \Omega $ from  \eqref{eq: ham after first rotation} more explicitly. 
We recall the partition of classical configurations in classes $\class$ defined in Section \ref{sec: Resonances: translation invariant systems}. 
The many-body Hilbert space can be decomposed accordingly: $\mathcal H = \bigoplus_{\class} \mathcal H_\class$. 
We also denote by $P_\class$ the projector on $\mathcal H_\class$, {$P_\class=\sum_{\eta \in \class} \str \eta \rangle \langle \eta \str$}; $\sum_\class P_\class$ is a partition of unity. 
Given a class $\class$, we denote by $F=F(\class)$  the set of frozen sites, and by $S = S(\class) \subset F^c(\class)$ the connected components of the complement $F^c(\class)$. 
It holds that $H_0' = \sum_\class P_\class H_0' P_\class$ and we write
\begin{multline*}
P_\class H_0' P_\class \; = \; P_\class \Big( \sum_x E_x^{(0)} \Big) P_\class \, + \, \sum_{S(\class)} H_{S(\class)} 
\\[1mm]
 \text{with } H_{S(\class)} \; = \; \coupling  \sum_{x:\,  \mathrm{supp}(U_{res, x}) \subset S (\class)} P_{\class} U_{res, x} P_{\class}
\end{multline*}
where we keep in mind that $H_{S(\class)} $ depends on $\class$ as well and where $\mathrm{supp}(O)$, for a local operator $O$, is the set of sites that $O$ acts on nontrivially. 
It is observed that, for any $\class$, the operator $P_\class (\sum_x E_x^{(0)}) P_\class$ is simply number when acting on states in $\class$, 
because in the first step, two states are declared resonant only if they have exactly the same uncoupled ($\coupling =0$) energy. 
Let us determine the eigenstates of $H'_0= \sum_{\class}  P_\class H'_0  P_\class $.  
For each class $\class$, they are products of classical configurations in $F$ and $H_S(\class)$ eigenstates $\alpha_S$ in the regions $S$, i.e.\@ 
\begin{equation}\label{eignestates afeter first step}
\Psi =  \eta_F \otimes  \mathop{\otimes}\limits_{S} \alpha_{S}
\end{equation}
Such an eigenstate is localized in most places, since we know from Section \ref{sec: example without percolation} that, for most states,  $\sum_S \str S \str \ll \str F \str$.
However, the nature of $\alpha_S$ plays a crucial role in the following.


\paragraph{Fusing two classes in second order: exemplary case.}
{As a first step to establish that all classes $\class$ will eventually be fused in the second step of the scheme, let us consider a class $\class$ where  there is at least one `big' (to be quantified later) component $S$, that we denote by $S^*$. This is of course typically the case as the volume grows large. 
We write $S^{*}=[a,b]$, we fix a certain $\eta_F$ and fractions $f_0,f_1,f_2 \in \mathbb{N}/\str S^{*} \str$ (with $f_0+f_1+f_2=1$) and we let $\class$ be the class of $\eta$ coinciding with the prescribed $\eta_F$ in $F$ and satisfying
\begin{equation} \label{eq: ballistic eigenstate}
\forall x \in S^{*}: \, \eta_{x}  \in \{0,1,2\}, \quad   \#\{x \in S^{*}:\eta_x =i\}=f_i \str S^{*} \str. 
\end{equation}
(and of course specifications for the other components $S$, which are completely irrelevant for this argument). }
The resonant Hamiltonian $H_{S^*}$ describes a system of interacting bosons. 
It seems safe to assume that this system is ergodic (there is no reason for it to be localized and it is not integrable in any obvious way). 

We show here that, in the second step, this class needs to be fused with the class $\class'$,
which is the same as $\class$ except that $\eta'_{a-1}=\eta_{a-1} -1$ and $f_i' \str S \str = f_i \str S \str +\delta_{i,1}$ (in particular $\class$ and $\class'$ have the same spatial structure, i.e.\@ the same set $F$ and the same sets $S$); we remark that we need  $\eta'_{a-1}\geq 5$, otherwise $a-1$  is not a frozen site and  $\class,\class'$ are not full classes.
The class $\class'$ is taken as an example: important is that one boson has been absorbed/ejected into/from $S^*$ and, for the moment, that $\class$ and $\class'$ have the same spatial strucure.
Eigenstates of $P_{\class'} H_0' P_{\class'}$ are denoted by 
\begin{equation*}
\Psi' =  \eta'_F \otimes  \mathop{\otimes}\limits_{S}\alpha'_{S}.
\end{equation*}


To establish that $\class$ and $\class'$ need to be fused in the second step, we show below that,
provided that $\str S^* \str$ is large enough,  for any eigenstate $\Psi$ of $P_{\class} H_0' P_{\class}$, there are many eigenstates $\Psi'$ of $P_{\class'} H_0' P_{\class'}$ resonant with $\Psi$, i.e.\@ such that they satisfy \eqref{eq: condition for fusing}: 
\beq \label{eq: condition for fusing repeat}
\str \langle \Psi \str \coupling^2 U' \str   \Psi' \rangle \str  \; \gg \; \str \langle \Psi \str H_0' \str   \Psi \rangle - \langle \Psi' \str H_0' \str   \Psi' \rangle \str. 
\eeq
Indeed, according to the definition of classes given in Section \ref{sec: Resonances: translation invariant systems}, this inequality indeed implies that the classes $\class$ and $\class'$ have to be fused.

\paragraph{Proof of  \eqref{eq: condition for fusing}.}
The main point to show \eqref{eq: condition for fusing} is this. 
Let us write $\mathcal H_\class = \mathcal H_{F(\class)} \otimes {\otimes}_{S(\class)} \mathcal H_{S(\class)}$, and let us denote by $\dd_{S^*}$ the dimension of $\mathcal H_{S^*(\class)}$ (we could equally well define $\dd_{S^*}$ as the dimension of $\mathcal H_{S^*(\class')}$ as these two quantities are of the same order).
We will show that, while the l.h.s.\@ of \eqref{eq: condition for fusing} behaves like $1/\sqrt{\dd_{S^*}}$, for given $\Psi$ one typically finds $\Psi'$ such that the r.h.s.\@ behaves like $1/\dd_{S^*}$, so that \eqref{eq: condition for fusing} holds for a large enough spot $S^*$.

Let us first  estimate
\begin{equation}\label{eq: overlap}
\langle \Psi \str \coupling^2 U' \str   \Psi' \rangle
\end{equation}
It is an easy check that all (low order in $\coupling$) contributions to $P_{\class} \coupling^2 U' P_{\class'}$ are located near site $a$ and the simplest contribution (namely, consisting of only two creation/annihilation operators) of order $\coupling^2$ is
$$
 \coupling^2 P_{\class}  b^*_{a-1}b_{a}  P_{\class'}.
$$
so that  \eqref{eq: overlap}  is, up to higher orders
$$
\coupling^2  \langle \eta_{F} \str b^*_{a-1} \eta'_{F} \rangle  \,\,  \langle \alpha_{S^*}\str b_{a}  \alpha'_{S^*}\rangle
$$
The first factor is of order $\sqrt{\NumLevels}$ by our choice of $\eta_F,\eta'_F$. The squared modulus of the second factor is the expectation value of 
$$
\langle \alpha_{S^*}\str P \str   \alpha_{S^*}\rangle \qquad \text{with}\qquad  P= \str  b_{a} \alpha'_{S^*}\rangle \langle b_{a} \alpha'_{S^*} \str .
$$ 
To estimate this, we invoke the ETH (Eigenstate Thermalization Hypothesis, see \cite{rigolnature} and references therein) claiming that the ensemble defined by just one eigenvector is in a certain sense equivalent to {an equilibrium ensemble at the appropriate values of the conserved quantities. In this case, we take as equilibrium ensemble the uniform (i.e.\ tracial) state on $\caH_{S^*}$ This yields 
$$
\langle \alpha_{S^*}\str P \str   \alpha_{S^*}\rangle 
\; \sim \;   \frac{1}{\dim(\caH_{S^*})}   \mathrm{Tr}_{\caH_{S^*}} ( P) 
\, = \,  \frac{1}{\mathrm{d}_{S^*}} \, = \, \mathrm{e}^{-s \str S^* \str}
$$
where $s$ is the corresponding entropy density.  }
It remains to estimate the right-hand side of \eqref{eq: condition for fusing}: If we choose $\Psi'$ so as to minimize this side, then it is of the order of the level spacing, which is
$$
W \frac{1}{\mathrm{d}_{S^*}} \sim e_0\str S \str e^{-s \str S^* \str}
$$
with $W \sim e_0\str S^* \str $ the width of the spectrum and $e_0$ the energy density.  Hence \eqref{eq: condition for fusing} reads
$$
\coupling^2  \sqrt{\NumLevels}   e^{-s \str S^* \str/2}  \geq  e_0 \str S^* \str  e^{-s \str S^* \str}
$$
which is satisfied provided that, roughly, 
$$
\str S^* \str \geq  C \big| \log(\coupling  \NumLevels^{3/4}) \big| , 
$$ 
where we also used that $ e_0 \sim 1/\NumLevels$ since the $\caE(1), \caE(2) \sim 1/\NumLevels$, and we simply wrote $C$ for parameters of order $1$. 
The reason that we call this argument `non-perturbative' is of course that the size of spots that we need to consider here grows as $\coupling \to 0$ (recall that $\coupling \NumLevels \ll 1 $), unlike in the  examples of Section \ref{sec: example with percolation}.

\paragraph{Generalization: fusing almost all classes.}
Up to now, as our conclusion did not depend on the value of  $\eta_{a-1}$ (provided that $\eta_{a-1} \ge 6$), we have shown that localization at site $a-1$ is completely loss due to the ergodic spot $S^*$.
We can now generalize our argument to show that the class $\class$ (with $\eta_{a-1}=6$) needs also to be fused with classes $\class''$ having a component $\{a-1\} \cup S^*=[ a-1,b]$. 
By translation invariance, these classes have then in turn to be fused with classes with a component $[ a-1, b-1]$, 
allowing us to establish that the ergodic spot initially located in $S^*$ can move accross the chain. 
Clearly, by the same mechanism, the spot can also carry bosons from any place to any other one, so that the environment can be modified. 


Let us now see how to connect $\class$ with $\class''$. 
Instead of considering  {(for clarity, we omit here all the product over all $S\neq S^*$, since $\alpha_S$ is the same for all vectors involved)}
$$
\Psi =  \eta_F \otimes  \alpha'_{S^*}, \qquad    \Psi'=  \eta'_F \otimes   \alpha'_{S^*} 
$$
as above, we consider 
$$
\Psi =  \eta_{F} \otimes  \alpha'_{S^*}, \qquad     \Psi''=  \eta''_{F\setminus(a-1)} \otimes  \alpha''_{\{ a-1\} \cup S^*} 
$$
with $\eta''_{F\setminus(a-1)}$ equal to the restriction of $ \eta_{F}$ to $F\setminus(a-1)$. 
The overlap \eqref{eq: overlap} is now calculated as 
\beq \label{eq: overlap new}
\str \langle \Psi \str \coupling^2 U' \str   \Psi'' \rangle\str  =  C\coupling^2   (\langle \Psi'' \str P \str   \Psi'' \rangle)^{1/2}
\eeq
with $P$  the one-dimensional projector with range $ \coupling^2 U' \Psi$. 
The factor $\sqrt{\NumLevels}$ is now missing (it is replaced by $C$ as $\eta_{a-1}$ is necessarily small (in fact, equal to $4$) in order that one application of the $U'$-term can enlist this site into the bubble.

\paragraph{Application to the  first model of Section \ref{sec: example without percolation}?}
For that first model, the class specified in \eqref{eq: ballistic eigenstate} would, for example, only have $\eta_x \in \{0,1\}$. Then $H_S(\class)$ would describe a system of one-dimensional hard core bosons with nearest neighbor hopping, which is integrable. This would have invalidated the assumption of ETH. However, even if one would have concluded that after the second iteration step, the Hamiltonian is localized, then one can continue the procedure and at some step the resonant Hamiltonians in the delocalized regions $S$ would generically be ergodic\footnote{Of course, this relies on the folk belief that ergodicity is generic.}  because, when sufficiently many perturbation terms are included, the range of the hopping and the boson-boson interaction grows. 
Hence, all what was really necessary for our argument is that, everywhere in space, there are subspaces in Hilbert space in which the system is ergodic.

\end{document}